\documentclass[namedreferences,hyperref,optionalrh]{spr-sola}
\usepackage{graphicx}        
\usepackage{color}           

\usepackage{newtxtext}
\usepackage{newtxmath}
\usepackage[T1]{fontenc}
\usepackage{subfig}
\usepackage{booktabs}
\usepackage{multicol,multirow}
\usepackage{hhline}
\usepackage{siunitx}
\usepackage{fmtcount}
\usepackage[version=4]{mhchem}
\usepackage{csquotes}
\MakeOuterQuote{"}
\usepackage{gensymb} 
\usepackage{array}
\usepackage{calc}
\newcolumntype{M}[1]{>{\centering\arraybackslash}m{#1}}

\chardef\us=`\_

\begin{document}

\begin{frontmatter}
\title{Brightenings AnD Polarity Inversion Tracking (BADPIT) method for studying solar active region evolution before major solar flares}

\author[addressref={aff1,aff2},corref,email={a.andre-hoffmann@uoi.gr}]{\inits{A.}\fnm{Augustin}~\snm{André-Hoffmann}\orcid{0009-0005-2036-4199}}
\author[addressref={aff3,aff4,aff5}]{\inits{M.B.}\fnm{Marianna~B.}~\snm{Korsós}\orcid{0000-0002-0049-4798}}
\author[addressref=aff1]{\inits{A.}\fnm{Alexander}~\snm{Nindos}\orcid{0000-0003-0475-2886}}
\author[addressref=aff1]{\inits{S.}\fnm{Spiros}~\snm{Patsourakos}\orcid{0000-0003-3345-9697}}
\author[addressref={aff6,aff7}]{\inits{M.K.}\fnm{Manolis~K.}~\snm{Georgoulis}\orcid{0000-0001-6913-1330}}
\author[addressref={aff2,aff4,aff5}]{\inits{R.}\fnm{Robertus}~\snm{Erdélyi}\orcid{0000-0003-3439-4127}}
\address[id=aff1]{Section of Astrogeophysics, Department of Physics, University of Ioannina, 45110 Ioannina, Greece}
\address[id=aff2]{Solar Physics and Space Plasma Research Centre (SP2RC), School of Mathematical and Physical Sciences, University of Sheffield, Hicks Building, Hounsfield Road, Sheffield, S3 7RH, UK}
\address[id=aff3]{School of Electrical and Electronic Engineering, University of Sheffield, Amy Johnson Building, Portabello Street, Sheffield, S1 3JD, UK}
\address[id=aff4]{Department of Astronomy, E\"otv\"os Lor\'and University, P\'azm\'any P\'eter s\'et\'any 1/A, H-1117 Budapest, Hungary}
\address[id=aff5]{Gyula Bay Zolt\'an Solar Observatory (GSO), Hungarian Solar Physics Foundation (HSPF), Pet\H{o}fi t\'er 3, H-5700 Gyula, Hungary}
\address[id=aff6]{Space Exploration Sector, Johns Hopkins Applied Physics Laboratory, Laurel, MD 20723, USA}
\address[id=aff7]{Research Center for Astronomy and Applied Mathematics, Academy of Athens, 11527 Athens, Greece (on leave of absence)}

\runningauthor{A. André-Hoffmann et al.}
\runningtitle{BADPIT method for studying pre-flare solar AR activity}

\begin{abstract}
This study investigates the relationship between extreme ultraviolet (EUV) transient brightenings (TBs) and the onset of GOES X-class solar flares in active regions (ARs).
We introduce the Brightenings AnD Polarity Inversion Tracking (BADPIT) method that can detect TBs across multiple SDO/AIA channels.
To identify TBs, we impose two independent thresholds: a 3-$\sigma$ intensity-based criterion and a power law divergence approach.
We demonstrate the application of BADPIT through a flaring and a non-flaring AR for 24 hours as a pathfinder to a comprehensive statistical study for a complete performance verification: the studied ARs are the non-flaring AR 13186 and the flaring AR 11429, both sharing a similar Hale sunspot classification.
Preliminary results are encouraging: significantly more TBs are detected in the flaring AR, with up to five times more 3-$\sigma$ thresholded TBs identified, while power law thresholded events were frequent only in the flaring AR and mostly absent in the non-flaring AR. 
In the two ARs that we studied both the power law threshold method and the 3-$\sigma$ method show a potential to act as diagnostic tools for distinguishing between imminent flaring and non-flaring conditions several hours before the onset of major solar flares.
However, our work is a proof-of-concept study, given the limited number of ARs we studied; its reliability as a forecasting tool will be investigated in a follow-up study in which a large sample of ARs will be analysed.

\end{abstract}
\end{frontmatter}

\section{Introduction}\label{sec:intro}

Many properties of solar active regions have been deemed important in describing the pre-flare state of flaring regions, in par with significant progress in predicting major solar flares in the Solar Dynamics Observatory (SDO) era \cite[see, for more details,][]{Georgoulis2024}.
The SDO mission has provided continuous observations of the solar atmosphere through the Atmospheric Imaging Assembly (AIA; \citealt{Lemen2012}) and of the photospheric magnetic field through the Helioseismic and Magnetic Imager (HMI; \citealt{Scherrer2012}).

Most studies aiming to identify signatures of imminent flaring in ARs rely on the evolution of magnetic fields \citep[see e.g.][ and references therein]{Canfield1975, LekaBarnes2003, Florios2018, Korsos2019, Korsos2024, Soos2022,Kontogiannis2023}.
In particular, strong and sheared polarity inversion lines (PILs) have been statistically linked to flares and coronal mass ejections (CMEs) \citep[see, e.g.][ and references therein]{Mikic1994, Jacobs2006, Sterling2004,schrijver_characteristic_2007, Threlfall2018,  Kontogiannis2018, Ji2023, biswal_2024}.
The PIL is a key region for the storage and release of magnetic free energy.
Strong-field, highly sheared, and extended PILs are closely associated with resistive (i.e., via magnetic reconnection) formation of magnetic flux ropes that will subsequently erupt \citep[see, for example][]{Green2018, patsourakos_decoding_2020}.
The build-up of magnetic shear and electric currents along the PIL makes it a favourable site for magnetic reconnection \citep{Priest2014,Temmer2021}, and two dominant mechanisms have been suggested as fundamental drivers of eruptive activity: magnetic reconnection and ideal magnetohydrodynamic instabilities \citep{Green2018, zhong_2023}.

To further improve the identification of the pre-flare/pre-eruptive phase in active regions, as well as refine existing flare prediction methods \citep[see the multiple flare predictors presented in][and references therein]{georgoulis_2021}, an important complement to magnetic-field-based precursors is provided by ultraviolet \citep{lu_2024} and extreme ultraviolet (EUV) observations \citep{bonte_2013}. 
Several studies have addressed EUV small-scale brightening prior to flares \citep[see e.g.][ and references therein]{Sterling2001, gopalswamy_coronal_2005, Kniezewski2024, Krista2025a, Krista2025b}, while others have investigated the occurrence of coronal dimmings \citep[see e.g.][ and references therein]{Harra2013, Kerr2021}.
Furthermore, growing evidence suggests that eruptive flares may be initiated by internal magnetic reconnection between a small-scale flare-triggering field and the overlying sheared arcade or magnetic flux rope.
These small-scale reconnection events, often manifesting as lower solar atmospheric pre-flare brightenings, are thought to act as triggers for flares \citep[e.g.,][]{Moore2001, Chifor2007, Kusano2012, Bamba2013, Chen2016, Bamba2017, Kusano2020, massa_efficient_2022, Krista2021, Leka2023,Dissauer2023, dissauer_brightenings_2025,  Krista2025a}.
Alternatively, as suggested by \citet{Young2018}, observed brightenings may simply reflect general photospheric and chromospheric dynamics and restructuring, such as emerging active regions, bald patches, sunspot moats, or light bridges, and may not be directly linked to subsequent energetic events.

In recent years, an increasing number of machine- and deep-learning studies have focused squarely on flare prediction.
These studies have leveraged the large-scale, high-cadence datasets of SDO/AIA \citep{Nishizuka2017, Jonas2018, Alipour2019, Woods2021, Leka2023, Sun2023}.
However, the AIA data used in these models is subject to significant spatial degradation over time \citep[see ][]{dos_santos_multichannel_2021}, with this downscaling partly hindering the identification of small-scale features that may be critical to understanding the underlying flare initiation mechanisms.
In summary, observational and theoretical/modelling work suggests that magnetic reconnection may play a key role in forming pre-eruptive magnetic configurations and triggering flares, with transient brightenings (TBs) serving as a "smoking gun" of these reconnections events.

\citet{Chifor2007} reported transient precursor brightenings around PILs between 2 and 50 min before a flare's impulsive phase.
Work such as this motivate us to investigate the evolution of brightenings in such regions over an extended, 24-hour interval prior to major flaring.
Our objective is to develop a method to determine whether there are systematic differences in the low/intermediate-activity between non-flaring (no flares) and flaring (X-class flares) ARs based on the characteristics of their TBs and how are the TB distributions correlated with the PIL area.

To enable brightening detection around the PIL, we introduce the Brightenings AnD Polarity Inversion Tracking (BADPIT) method, which identifies and characterises TBs using time-series analyses of HMI and AIA observations.
In our study the TBs could include activity ranging from nanoflares, microflares, and subflares up to major flares.
This is done due to the way we detect them (see section~\ref{bright det}) and because it is beyond the scope of our work to determine their energy budgets.
We aim to study the pre-flare evolution of active regions over a 24-hour period and further investigate whether the evolution of TBs near the PIL is connected to upcoming energetic events.
The proposed methodology follows a similar hypothesis as recently put forward by \citet{dissauer_brightenings_2025}.
They analysed 2-hour intervals during 3 intervals: the flaring interval and the pre- and post-event associated intervals.
They suggested that, prior to flaring, TBs form large clusters near the future flare site, concentrated around strong-gradient PILs.
In contrast to their topological approach, our work focuses on quantifying the evolution of the TB population and tracking its long-term behaviour over a substantial observational period.

This work is structured as follows: In Section~\ref{sec:data}, we introduce the dataset used in our analysis.
Section~\ref{methodo} describes the key components of the BADPIT method for identifying small-scale TBs.
Section~\ref{sec:test} shows the results obtained with the method on a test pair of ARs exhibiting flare activity levels typical of flaring (period prior an X-class flare) and non-flaring (no C-class or above flare) ARs.
Finally, Section~\ref{sec:discu} presents the conclusions and outlines some future directions of research.

\section{Data}\label{sec:data}

We use cutouts from the full-disk line-of-sight (LoS) photospheric magnetograms from SDO/HMI \citep{Scherrer2012}, along with co-aligned SDO/AIA \citep{Lemen2012} Level 1.5 EUV images in multiple narrow-band channels centered at 94, 131, 171, 193, and 1600\,\AA.
To avoid projection effects on the LoS magnetograms, the locations of the two studied ARs are within $-45\degree$ and $+45\degree$ of central meridian distance.
Additionally, for the SDO/AIA images and prior to the start of the analysis, we applied a desaturation procedure described in Section~\ref{sec:desat} and normalised the data by the exposure time.

For each target active region (see details in Table~\ref{tab:datasets}), cutouts were extracted from both the AIA images and the corresponding HMI magnetograms using Stanford University's JSOC export service\footnote{Procedure for obtaining SDO cutouts from the JSOC service by Dr Peter R Young: \href{https://www.pyoung.org/quick_guides/aia_eis_wiki.html}{https://www.pyoung.org/quick\_guides/aia\_eis\_wiki.html}}).
These multi-layer, co-temporal datasets provide a comprehensive view of the evolution of the ARs from the photosphere to the corona.

The 94\,\AA\ channel observes plasma with temperatures from several hundred thousand Kelvin ($\sim$0.6\,MK) to around 5-6\,MK, particularly during flaring activity.
This channel is especially sensitive to emission from regions on the verge of flaring. 
The 131\,\AA\ wavelength provides insight into both high-temperature flaring plasma ($\sim$10\,MK) and cooler coronal structures ($\sim$0.4\,MK), making it useful for studying both dynamic and quiescent phases. 
The 171\,\AA\ channel is ideal for tracing the structure of coronal loops at approximately 0.6\,MK, while 193\,\AA\ captures hotter, ambient coronal plasma around 1.5\,MK and is useful for examining coronal holes and hotter flare plasmas. 
Finally, the 1600\,\AA\ channel observes emissions from the upper photosphere up to below the transition region, capturing lower-atmosphere dynamics in the range of approximately $0.01-0.1$\,MK. 

By combining observations from HMI magnetograms with observations from multiple AIA channels covering atmospheric layers from the photosphere to the corona, we are able to track the evolution of TBs across a broad range of heights and temperatures, at the same time monitoring the magnetic activity underneath them. 
This multi-channel approach enables us to identify the optimal combination of wavelengths for detecting TBs in the vicinity of the PIL.

\begin{table*}
    \caption{Information about NOAA ARs 13186 and 11429 used in this study, including observational parameters from AIA and HMI instruments. All times shown are UT.
   The observation periods for both ARs are also provided.}
    \label{tab:datasets}
    \centering
    \begin{tabular}{
    | >{\raggedright\arraybackslash}p{0.25\linewidth}
    || >{\centering\arraybackslash}p{0.3\linewidth}
    | >{\centering\arraybackslash}p{0.3\linewidth} |
    }
        \hhline{|t:===:t|}
        \multicolumn{3}{||c||}{Datasets} \\
        \hhline{|b:=:t:==:b|}
        Instrument & AIA & HMI\\
        \hhline{|-||-|-|}
        Wavelength (in \AA) & 94, 131, 171, 193 and 1600 & 6173 (line-of-sight magnetograms)\\
        Pixel size & $0,6''$ & $0,5''$\\
        Cadence & 12~s (24~s for the 1600~\AA) & 12~min\\
        \hhline{|=::==|}
        AR number & \multicolumn{2}{c|}{11429}\\
        Stage & \multicolumn{2}{c|}{End of emergence}\\
        Hale class & \multicolumn{2}{c|}{$\beta\gamma\delta$}\\
        Flaring (M- or X- class) & \multicolumn{2}{c|}{Yes}\\
        Eruptive & \multicolumn{2}{c|}{Yes}\\
        Time of observation & \multicolumn{2}{c|}{2012-03-06T00:00 -- 2012-03-07T00:00}\\
        Position & \multicolumn{2}{c|}{N17E27 at 2012-03-06T00:00}\\
        \hhline{|-||-|-|}
        AR number & \multicolumn{2}{c|}{13186}\\
        Stage & \multicolumn{2}{c|}{Fully developed}\\
        Hale class & \multicolumn{2}{c|}{$\beta\gamma\delta$}\\
        Flaring (M- or X- class) & \multicolumn{2}{c|}{No}\\
        Eruptive & \multicolumn{2}{c|}{No}\\
        Time of observation & \multicolumn{2}{c|}{2023-01-16T00:00 -- 2023-01-17T00:00}\\
        Position & \multicolumn{2}{c|}{N25W15 at 2023-01-16T18:58}\\
        \hhline{|-|b|--|}
    \end{tabular}
\end{table*}

\section{Methodology}\label{methodo}

The BADPIT method tracks changes in TB activity using AIA observations over a continuous 24-hour period.
The spatial location of the detected TBs is also compared with the polarity inversion area (PIA), namely the PIL and its surroundings.
In a distinct way than that of the  \citet{dissauer_brightenings_2025} study, we quantify and track the evolution of TBs and the TB population over selected 24-hour intervals.

To demonstrate the application of the BADPIT method, we have selected two active regions, namely NOAA ARs 13186 and 11429.
In this work, the two ARs serve as representatives of non-flaring (AR 13186) and flaring (AR 11429) active regions.
The first hosts no flares above GOES C1.0 during the interval of interest, as well as in the preceding and following 48 hours.
AR 13186, observed in mid-January 2023 (see Table~\ref{tab:datasets} for detailed information), is estimated to be fully developed.
Remarkably, however, it has a Hale-class ($\beta\gamma\delta$) typically associated with strong magnetic activity which corresponds to ARs with bipolar sunspot groups accompanied by minor spots of opposite polarity and with at least two opposite polarity umbrae sharing one penumbrae \citep{hale_magnetic_1919, kunzel_zur_1965}.
The second hosts GOES C- and M-class flares during the interval of interest followed by two X-class flares over a period of 1 hour. 
AR 11429, observed in early March 2012 (see Table~\ref{tab:datasets}), is incompatible with Hale’s polarity law and exhibits a conspicuous $\delta$-spot magnetic configuration, indicating a complex magnetic structure likely to produce major flares and eruptions \citep{Georgoulis2019}.
At this moment, the AR is at the end of its emerging phase or reaching the beginning of its fully developed phase.

AR 11429 is the solar source of a few major space weather events.
Among multiple C- and M-class flares, this AR hosted the two X-class flares mentioned above and produces fast CMEs on March 7, 2012, with speeds exceeding 2000\,km/s \citep{Patsourakos2016}.
The events occurred within an hour of each other and starts right after the end of the interval we study, as previously mentioned.
Using multi-wavelength data from SDO and STEREO, alongside non-linear force-free magnetic field extrapolations, \citet{Chintzoglou2015} identified weakly twisted magnetic flux systems interpreted as pre-eruptive magnetic flux ropes.
These structures formed during flaring events hours before the eruptions and eventually evolved into the cores of the two CMEs.
\citet{Syntelis2016} revealed by a differential emission measure analysis the presence of hot plasma (\(\log T = 6.8 - 7.1\)) in both the eastern and western parts of AR 11429, where the CMEs originated from.
These magnetic flux ropes displayed gradual heating and upward motion before the eruptions.
Therefore, this highly active AR 11429 is an ideal candidate to the BADPIT method application to a flaring active region case.
Table~\ref{tab:datasets} summarises the flaring activity, the datasets used, and the study periods for the two ARs. 

In Sections \ref{sec:desat} -- \ref{analy:magneticfield} below, we present a step-by-step application of the BADPIT methodology, which allows us to track the evolution of TBs over a 24-hour period in greater detail than previously attempted.
This improvement is achieved by first correcting for the saturation effects in the AIA observations, which would otherwise hinder the detection of TBs.

\subsection{Desaturation tool}\label{sec:desat} 

\begin{figure*}
  \centering
  \subfloat[Original data]{\includegraphics[width = .3\textwidth]{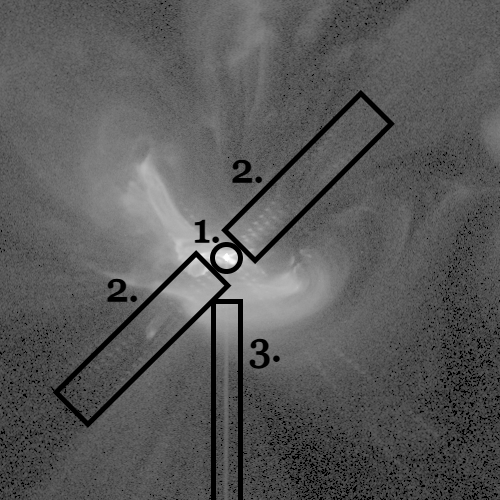}}
  \hspace{.1\textwidth}
  \subfloat[Post desaturation treatment]{\includegraphics[width = .3\textwidth]{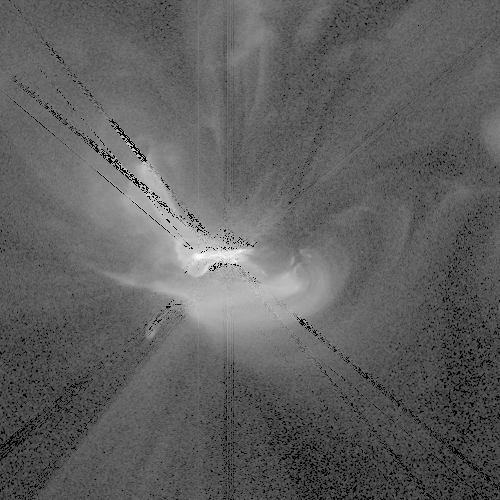}\label{fig:desat_result}}
  \caption{Example AIA 94\,\AA\ saturated frame (panel a) and its desaturated result (panel b) from AR 11429 observations on 06/03/2012 at 03:27:03UT.
  In panel A, the circle and boxes highlight examples of artefacts that must be removed before applying BADPIT: (1) saturated pixels, (2) diffraction arms, and (3) saturation trails.
  In panel B, remnants of the desaturation seen as dark diagonal lines can be observed but do not disrupt our procedure as they do not reach the detection threshold value.}
  \label{fig:sat}
\end{figure*}

A notable byproduct of this study is a new algorithm that was developed to achieve a sufficient --and appropriate for our purpose-- desaturation of the AIA images.
Desaturation is an essential step to ensure the quality and validity of the input data to the BADPIT method.

AIA images may be contaminated by artefacts originating from the design of the AIA telescope, which includes both an entrance filter mesh and a focal-plane filter mesh as part of its optical system.
These components are intended to block visible light and off-band EUV radiation but also introduce optical diffraction effects as illustrated in Fig.~\ref{fig:sat}.
Due to the mounting angles of the meshes (40\degree\ and 50\degree), characteristic diffraction patterns appear as four arms around saturated regions, oriented in the focal plane at 50\degree, 40\degree, $-40$\degree, and $-50$\degree\ \citep{grigis_aia_nodate}.
Saturation refers to pixels with intensities at or above 16000 $\text{DN}$ \citep[data number,][]{schwartz_desat_2015}.

In addition to saturated pixels, we account for two related types of artefacts: First, saturation trails, namely vertical streaks aligned along the detector’s $y$-axis, flagged to capture secondary effects from bright pixel blooming (see area 3 in Fig.~\ref{fig:sat}).
Second, diffraction arms, namely linear features aligned with axis defined by the telescope’s Point Spread Function (PSF), as described in the AIA PSF Characterisation and Image Deconvolution documentation \citep{grigis_aia_nodate}.
These diffraction patterns are flagged along their respective axes (see areas 2 in Fig.~\ref{fig:sat}).

To conservatively capture saturation effects, as well as saturation trails and diffraction patterns, and stay consistent with the approach by \citet{schwartz_desat_2015}, we flag all pixels with intensities above 12000 $\text{DN}$ as saturated (see area 1 in Fig.~\ref{fig:sat}).
Once all problematic pixels are identified in the EUV observations, they are replaced by values derived via linear interpolation in time.
Due to the way brightenings are detected (through thresholding methods), we add a sinusoidal component to the interpolated light curves.
This prevents the corrected points from triggering the brightening detection procedure (see below) and thus helps avoid the processing of false data.

\begin{figure}
    \centering
    \includegraphics[width=.6\linewidth]{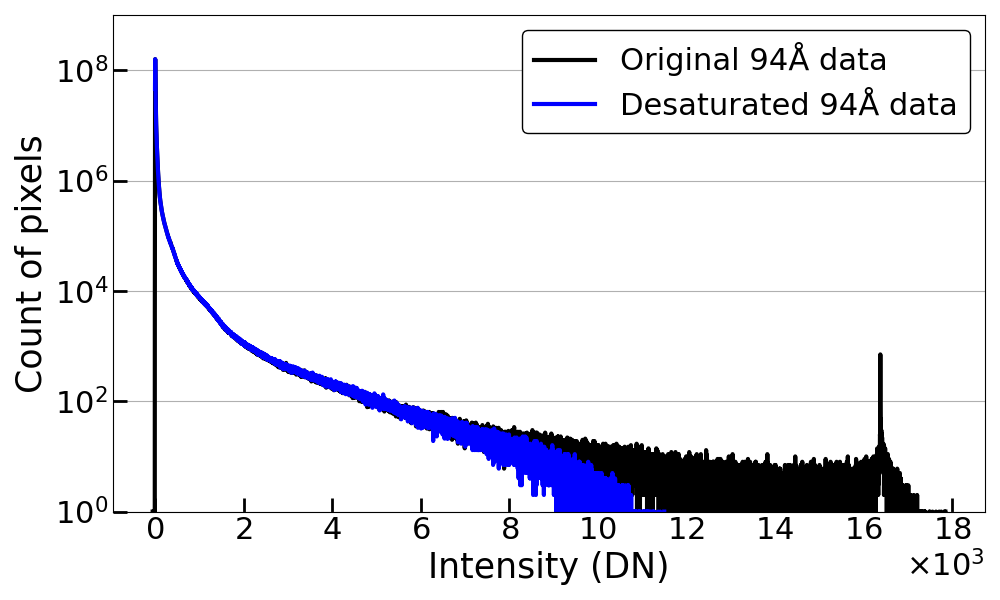}
    \caption{Comparison of the distribution function of all pixel's intensities (in $\text{DN}$) of the field-of-view from all frames over the 24-hour interval of interest of AR 11429, for the original 94\,\AA\ data (black) and the desaturated data (blue).}
    \label{fig:histogram sat-desat}
\end{figure}

The method described here will not have any effect on persistently bad pixels, even in cases where they would have a value over the saturation threshold.
Because they would not affect our study due to the detection criteria that will be presented in Section~\ref{bright det}, we do not try to detect and correct such artefacts.
Figure~\ref{fig:histogram sat-desat} shows the effect of desaturation on the original data.
Both the original (unprocessed) data and the desaturated ones have the same total count of pixels, but we observe a spike around $16000~\text{DN}$ in the original data (black curve) due to sensor saturation while this spike is absent in the desaturated data (blue curve) and correction has been successfully applied to most of the intensities above our saturation threshold.

Our desaturation procedure reduces the risk of false detections and ensures that the detected brightenings correspond to physically meaningful transient activity.
Nevertheless, caution is advised as the desaturation process itself can introduce new artefacts.
An example is shown in Figure~\ref{fig:desat_result}, where desaturation leads to dark patches along the PSF axis.
One might consider that we solved one problem only to generate another. 
However, these newly created artefacts are ignored by our brightening detection method due to the criteria we apply (explained in Section~\ref{bright det}).
Thus, they do not compromise our analysis.
Still, caution is invariably advised when employing desaturation methods: a careful investigation should always precede their application to evaluate their potential effects on the study.

BADPIT and our desaturation procedure were made available open-source in a Github repository\footnote{Github repository for the desaturation procedure and the BADPIT: \href{https://github.com/aandre-hoffmann/BADPIT}{https://github.com/aandre-hoffmann/BADPIT}}

\subsection{Brightening detection on the AIA observations}\label{bright det}

\begin{figure*}
    \centering
    \subfloat[]{\includegraphics[height=0.25\linewidth]{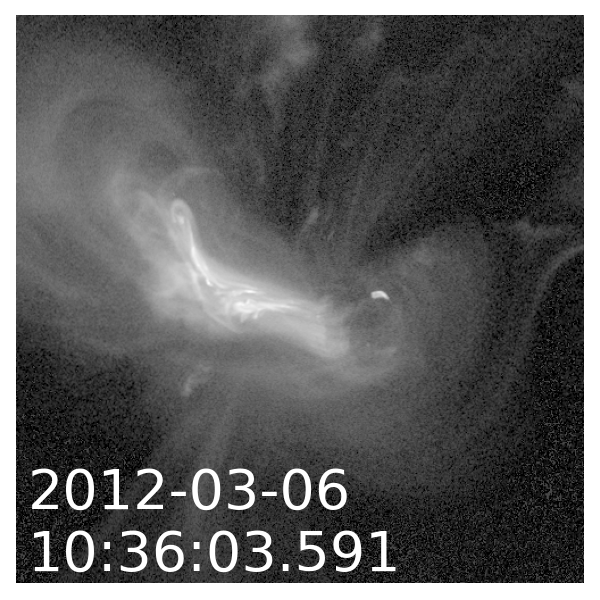}\label{diff a}}
    \subfloat[]{\includegraphics[height=0.25\linewidth]{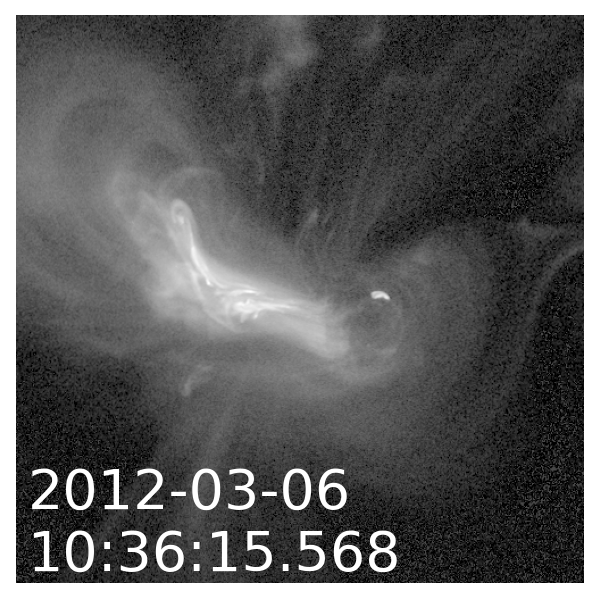}\label{diff c}}
    \subfloat[]{\includegraphics[height=0.25\linewidth]{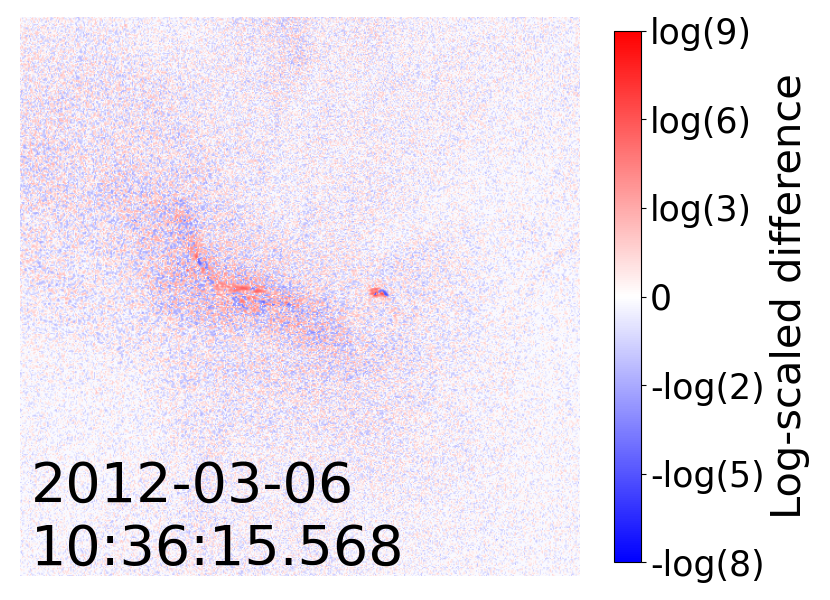}\label{diff b}}
    \caption{Resulting difference image (c) of two consecutive desaturated images (a and b). The two images used are AR 11429 AIA 94\AA\ images, 12 seconds apart.}
    \label{fig:diff bright}
\end{figure*}

After the desaturation process, brightenings in active regions can be identified.  
Brightenings are short-lived, localised events that are significantly brighter than their surrounding background.
In this paper, the term "transient brightenings" is used to describe a broad range of solar activity, from nano and microflares, up to major flares (see also section~\ref{sec:intro}).
They can also be visualised in sequences of difference images such as in Fig.~\ref{fig:diff bright}, where they appear as rapidly changing, coherent patches of pixels.

BADPIT detects intensity peaks in the light curves of each pixel within the field of view (FOV).
Brightenings are then defined as spatially coherent patches of high-intensity pixels.
To ensure that only physically meaningful events are detected and to minimise the likelihood of including artefacts into the analysis, we introduce two additional constraints, based on the technical limitations of the SDO/AIA observations (12-second cadence and $\sim 1.5''$ spatial resolution):

\begin{itemize}
  \item Spatial scale: Detected patches must span an area larger than 3 contiguous pixels (i.e., greater than $\num{5.67e5}~\si{km}^{2}$).
  \item Temporal scale: Detected events must persist for more than five consecutive frames (i.e., at least one minute).
\end{itemize}

These criteria restrict the analysis to brightenings that exhibit both spatial structure and sufficient temporal evolution to be confidently characterised as transient activity.
While smaller-scale events may indeed occur, the risk of ignoring them is deemed lower than the risk of including them in the analysis.
Moreover, difference images may reveal both real brightenings and background noise fluctuations, as shown in Fig.~\ref{fig:diff bright}.
By applying the spatial and temporal filters described above, we isolate only the relevant TBs.

Once we ensure that only the most statistically significant and morphologically consistent brightenings remain, we apply two methods to accurately identify brightenings' populations, namely:

\begin{itemize}
\item{\it 3-$\sigma$ threshold method}

This intensity threshold (3 times the standard deviation; 3-$\sigma$) is determined based on the Gaussian distribution of every pixels' intensity to remove the noise from our signal.
This 3-$\sigma$ threshold corresponds to a high-confidence criterion for detecting significant brightenings relative to the background \citep[see, for example][]{Nindos2020}.
It is used here to help us better understand small-scale transient activity and behaviour.
The threshold used for this detection method is pixel dependent, and as such, no pixel can continuously surpass its value.
However, in some situations, a pixel that would exceed the threshold once, or even at multiple occasions during the observed period, may not be considered a TB if it does not meet the additional spatial and temporal scale requirements previously mentioned.
Since this threshold is determined from the full-period distribution of intensities, it is not directly applicable to a potential future attempt of near real-time monitoring of TBs for forecasting purposes.

\item {\it power law fit based threshold method}

This threshold is globally defined, meaning it is not pixel-based.
It is spatially and temporally invariant for a given AR throughout the observation interval and is based on the intensity histogram of the AR over the entire period.

Our aim here is to use a power law fit as a tool to define a threshold for “strong” brightenings.
To achieve this, we set the threshold at the intensity where the histogram begins to deviate from the fitted distribution at high intensities.
We identify two configurations in the residuals, as shown in Figure~\ref{fig:fitting}, that serve as qualitative criteria for defining deviation values: a significantly increased spread in the intensity distribution (Fig.~\ref{fit 13186}) and a high-intensity frequency rollover (Fig.~\ref{fit 11429}).
This method allows us to identify brightenings that are substantially stronger than the background small-scale activity within the AR and that depart from an otherwise self-similar intensity distribution, represented by a power law.

Before adopting the power law fit, we compared the density distribution of the intensity histogram with other distributions frequently considered in similar studies, such as the log-normal or exponentially modified Gaussian (EMG) \citep[see ][]{dos_santos_multichannel_2021}.
The comparison is shown in Figures~\ref{fig:fitting} and \ref{fig:fit_multi}.
Overall, the intensity distribution appears to be better described by a power law ($y=ax^{-\alpha}$).
Nevertheless, fitting the intensity histogram with a power law is subject to several limitations.
Deviations from ideal power law behaviour at the lower end of the energy distribution may arise from instrumental limitations in both sensitivity and spatial/temporal resolution of the instrument.
With the decrease of event energies, it becomes increasingly difficult to distinguish genuine events from instrumental noise or background coronal emission, potentially causing an artificial turnover at the low-energy end of the distribution, making a straight power-law fit fail.
Furthermore, some of the weakest events may be driven by mechanisms different from those responsible for larger events, making a single power-law fit across a wide energy range potentially inappropriate.
\begin{figure*}
    \centering
    \subfloat[AR 13186]{\includegraphics[width=0.45\linewidth]{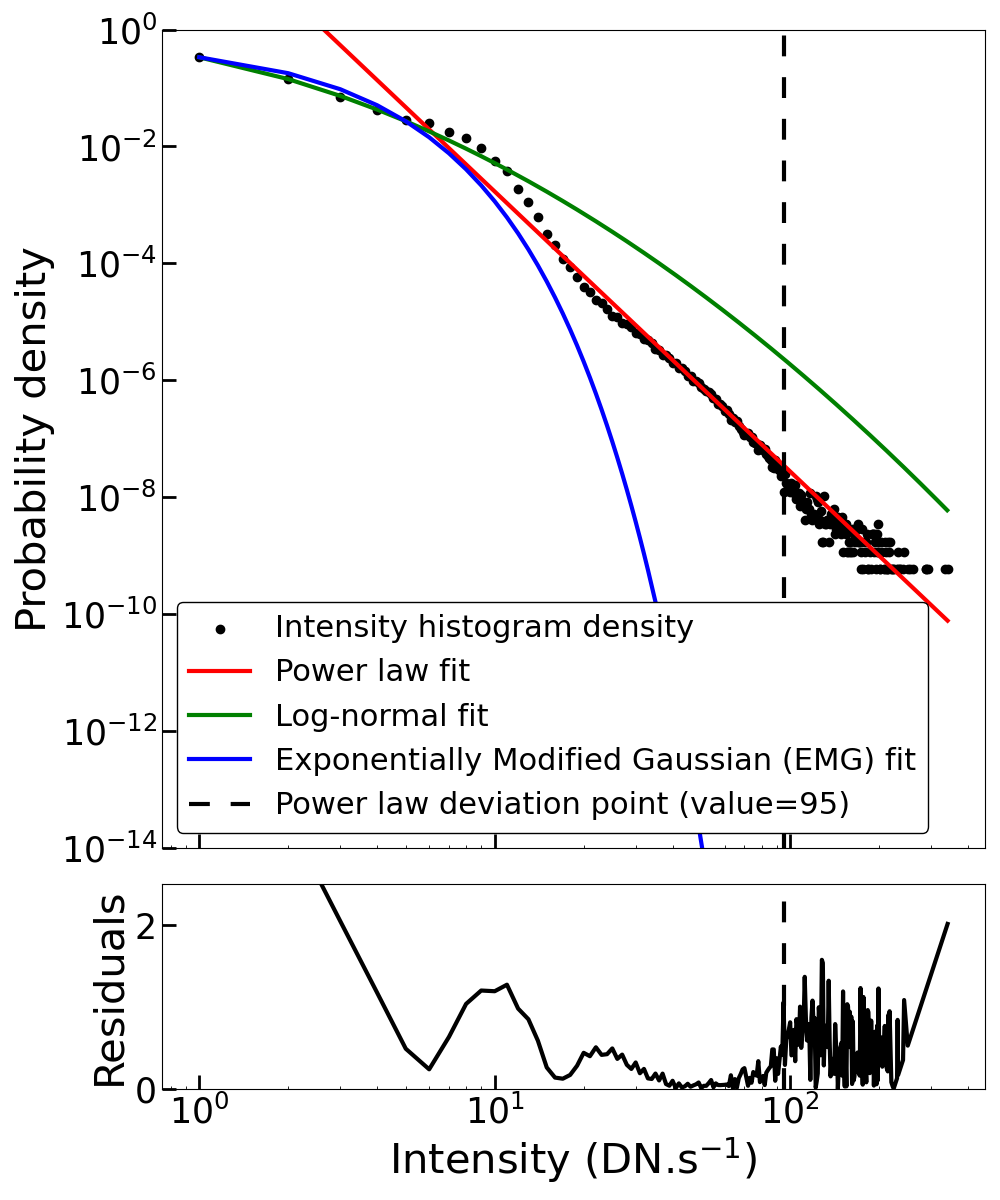}\label{fit 13186}}
    \subfloat[AR 11429]{\includegraphics[width=0.45\linewidth]{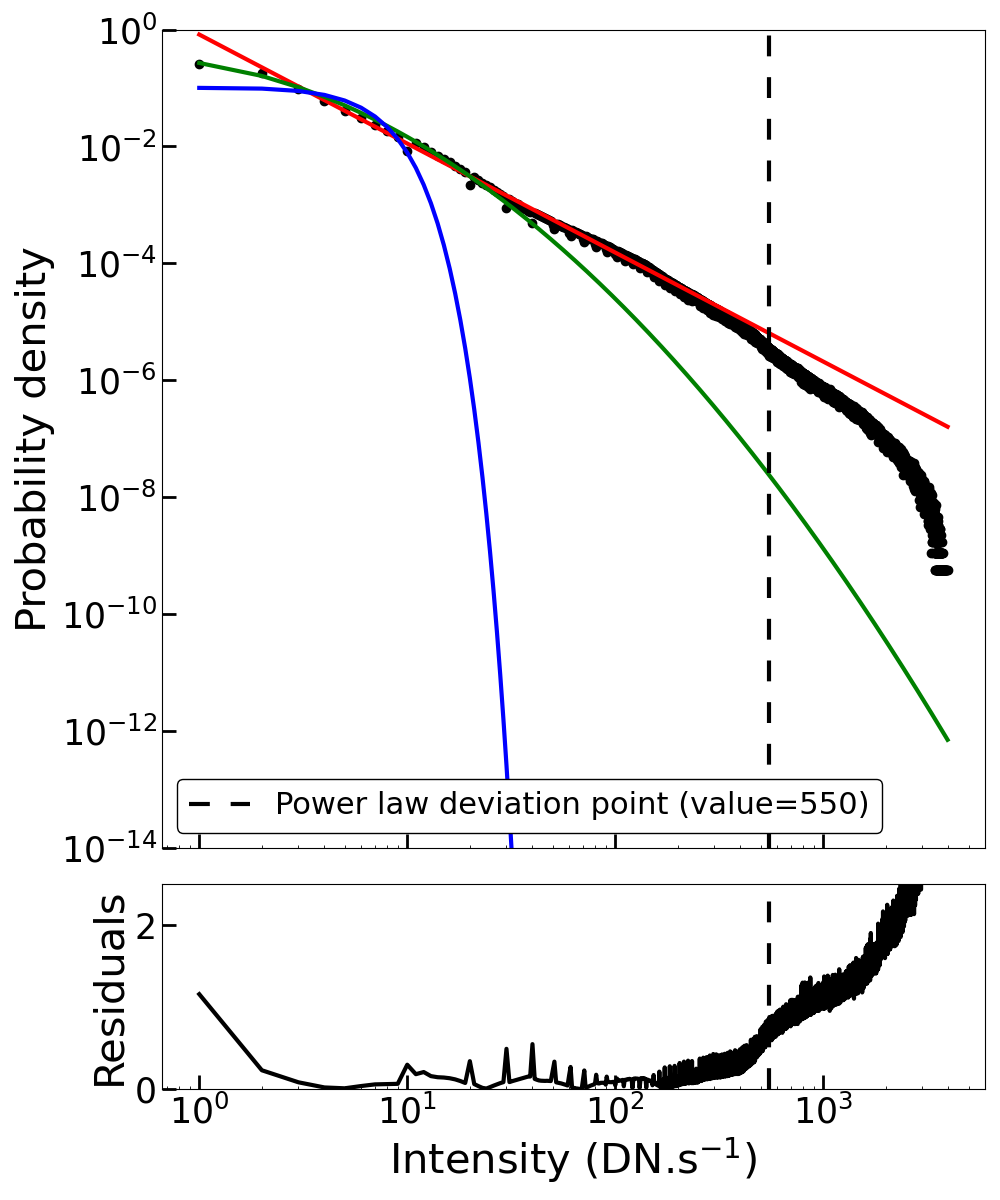}\label{fit 11429}}
    \caption{Probability density distribution of the 94\,\AA\ channel intensity (black curve) compared to different fitting methods, namely power law (red), log-normal (green), and exponentially modified Gaussian (blue).
    The exponent values of the power law fit are $4.80$ for AR 13186 (panel a) and $1.87$ for AR 11429 (panel b).
    The vertical dashed line represents the value at which the data separates from the power law fit.
    The bottom plots show the residuals from the power law fit to the histogram density in log-log space for each AR.}
    \label{fig:fitting}
\end{figure*}

Also, a clear difference in the power law index, $\alpha$, already provides qualitative information about the ARs: the non-flaring AR exhibits a steeper, “softer” distribution characterised by a large absolute $\alpha$ value, while the flaring AR displays a flatter, “harder” distribution with a smaller absolute $\alpha$ value.
This implies that intense, persistent emission occurs more frequently in the flaring AR, whereas the non-flaring AR is dominated by weaker emission patterns.
Consequently, the high-intensity tail of the distribution becomes more pronounced by the increase of activity.

Power-law behaviour can also be inferred at other wavelengths, although effective fitting ranges and power-law indices vary among the different EUV channels (see Fig.~\ref{fig:fit_multi}).
\begin{figure*}
    \centering
    \subfloat[AR 11429]{\includegraphics[width=\linewidth]{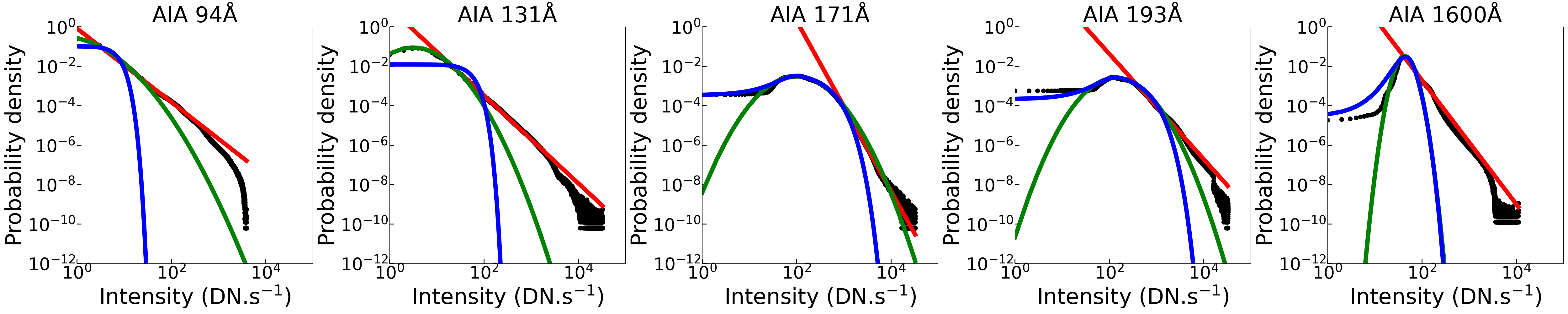}\label{multi fit 11429}}\\
    \subfloat[AR 13186]{\includegraphics[width=\linewidth]{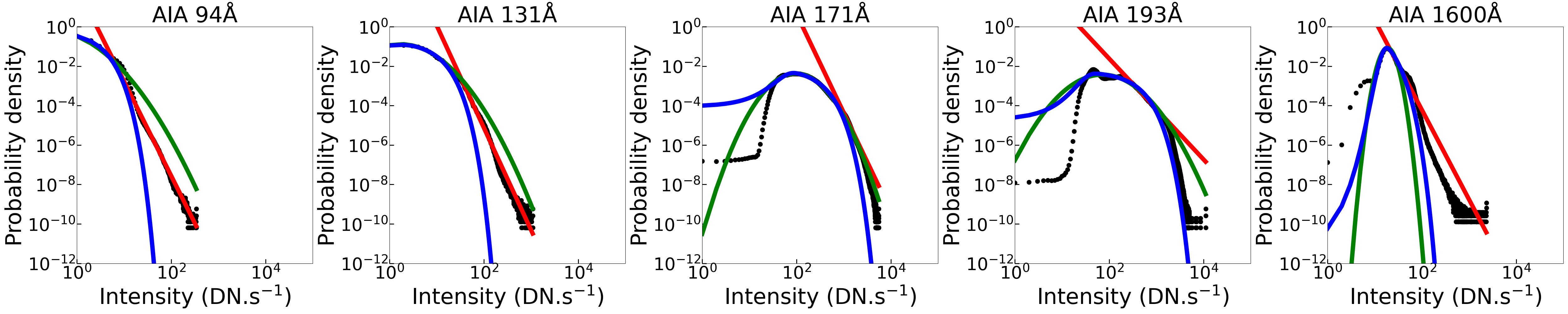}\label{multi fit 13186}}
    \caption{Probability density distribution of the intensity histograms (top line: flaring AR 11429; bottom line: non-flaring AR 13186) from multiple AIA channels (from left to right: 94\,\AA, 131\,\AA, 171\,\AA\, 193\,\AA\ and 1600\,\AA).
    The curves are colour-coded as in Fig.~\ref{fig:fitting}: black dots, red, green, and blue correspond to the probability density distribution, power-law, log-normal, and exponentially modified Gaussian, respectively.}
    \label{fig:fit_multi}
\end{figure*}
By adapting the power-law fit to each AIA channel, we obtain a corresponding TB detection threshold for all channels.
Due to different instrument sensitivity in every channel as well as differences in the observed plasma's temperature, it is expected to have different thresholds values, as shown in Table~\ref{tab:thresh}.
\begin{table*}
    \caption{Values in $\text{DN}\cdot\si{s}^{-1}$ of the power law based detection threshold for both AR 13186 and 11429 in multiple UV/EUV AIA channels.}
    \label{tab:thresh}
    \centering
    \begin{tabular}{c||c|c||}
        \hhline{~|t:==:t|}
          & \textbf{AR 13186} & \textbf{AR 11429} \\
        \hhline{|t:=::==:|}
        \multicolumn{1}{||c||}{\textbf{94\,\AA}} & 95 & 550 \\
        \multicolumn{1}{||c||}{\textbf{131\,\AA}} & 200 & 3000 \\
        \multicolumn{1}{||c||}{\textbf{171\,\AA}} & 2500 & 4000 \\
        \multicolumn{1}{||c||}{\textbf{193\,\AA}} & 1500 & 6000 \\
        \multicolumn{1}{||c||}{\textbf{1600\,\AA}} & 250 & 2500 \\
        \hhline{|b:=b::b==:b|}
    \end{tabular}
\end{table*}
Although there is ongoing debate regarding the best distribution to describe flaring-event probabilities is a power law or a log normal distribution \citep{verbeeck_2019}, the intensity histograms of the ARs examined in this study are generally better represented by a power law, regardless of their flaring activity.
The main exception is the 171\,\AA\ channel, which appears to be more consistent with a log-normal distribution.
It would be interesting to study this pattern on a larger dataset to have statistically relevant observations but this is not within the scope of our current study.
\end{itemize}

Utilizing the above, we found that the combination of the 3-$\sigma$ and power law threshold methods provides valuable insight into TB activity and effectively filters out both nearby quiet Sun regions and low-activity areas within the active region.
We will return to this discussion in Section \ref{sec:test}.
Distinct from the pixel-based 3-$\sigma$ method, the power law detection method captures generally stronger TBs, corresponding to the brightest TBs detected by the 3-$\sigma$ method.
The 3-$\sigma$ method captures a broader range of localised events, including both stronger and weaker events.
Therefore, the two threshold identification methods can be considered complementary.

\subsection{Analysis of Magnetic Field Observations}\label{analy:magneticfield}

\begin{figure}
	\centering
	\subfloat[HMI magnetogram of AR 11429.]{\includegraphics[width=0.4\linewidth]{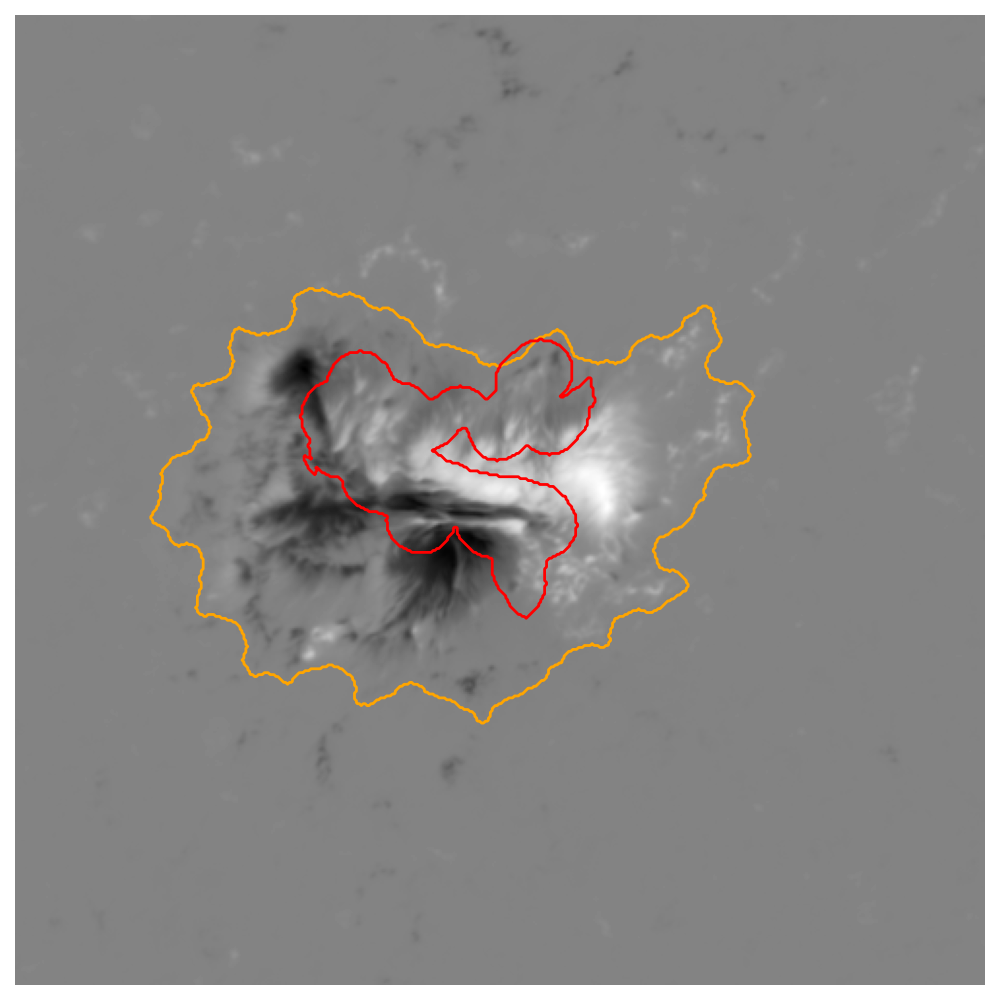}\label{fig:sdo_mag}}
	\hspace{0.05\linewidth}
	\subfloat[1600\AA\ AIA image of AR 11429.]{\includegraphics[width=0.4\linewidth]{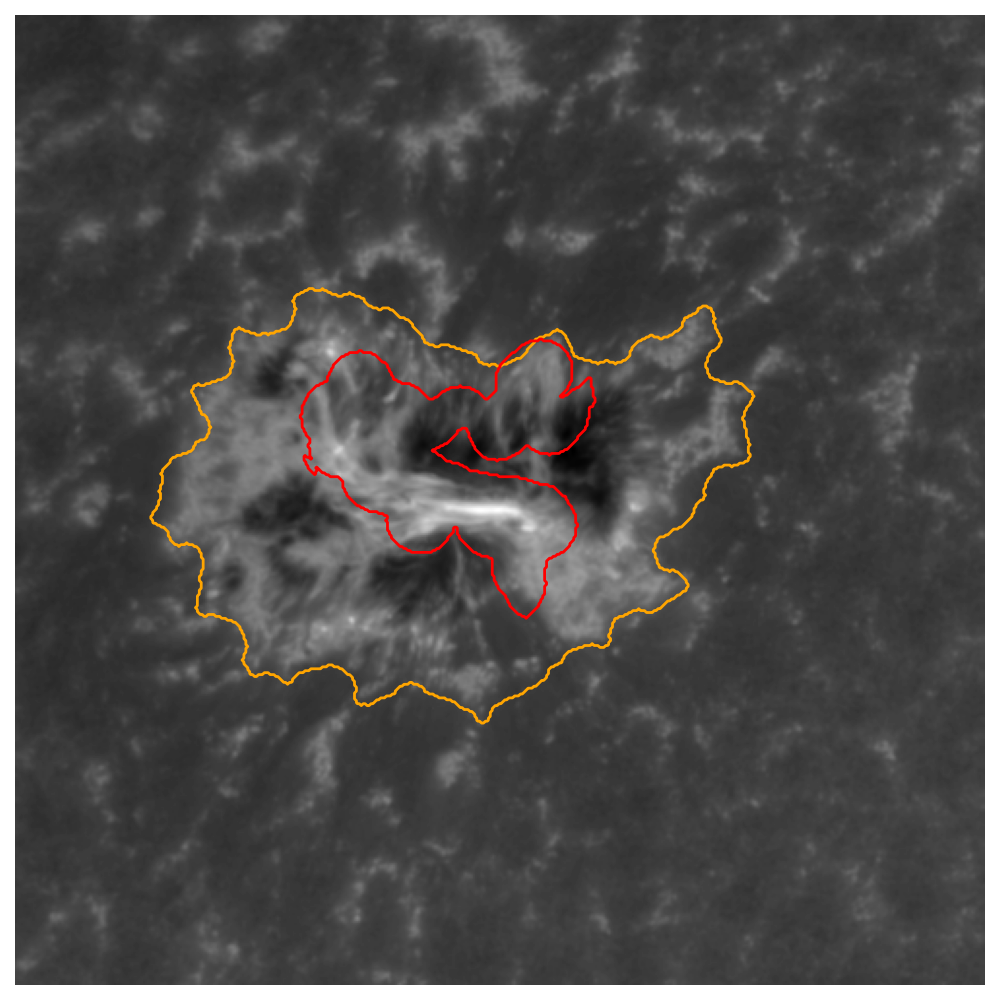}\label{fig:sdo_photosphere}}\\
	\caption{HMI magnetogram (a) and AIA 1600\,\AA\ (b) observations of AR 11429 averaged over a 3-hour period starting from 2012-03-06 at 09:00.
    The orange and red contours represent the AR area and the PIA, respectively.}
	\label{fig:masks}
\end{figure}

To examine the connection between the locations of TBs and the area of the PIL, we use LoS magnetic field data to generate masks around the PILs, as in the case of AR 11429 shown in Fig.~\ref{fig:sdo_mag} (magnetogram) and Fig.~\ref{fig:sdo_photosphere} (1600\,\AA), following the methodology presented in \citet{schrijver_characteristic_2007}.
LoS magnetograms are sufficient here since we restrict our analysis to ARs within central meridian distances of $\pm 45\degree$ as mentioned in Section~\ref{sec:data}. 

We define the PIA as the region which encompasses the PIL and extends into its immediate surroundings by $8''$ (16 SDO/HMI pixels).
The value was chosen empirically to allow a good overlap between both polarities of the AR magnetic field after multiple tests.
The broader PIA provides a useful indication of where magnetic reconnection associated with the PIL takes place, and thus, where TBs are most likely to occur.
The PIA detection algorithm operates by generating two masks based on where the absolute value of the magnetic field exceeds a preset threshold of $\pm 150$~G \citep{schrijver_characteristic_2007}.
These masks are then dilated by 16 pixels ($8''$), and the regions where the two expanded masks overlap identify areas where opposite polarities are sufficiently close to each other, thereby defining the PIA.
An example of the resulting PIA mask is shown for AR 11429 in Figure~\ref{fig:masks} and represented by the red contour over the LoS magnetic field (Fig.~\ref{fig:sdo_mag}) and the 1600\,\AA\ image (Fig.~\ref{fig:sdo_photosphere}).
Additionally, another mask, that of the entire AR, is also derived from HMI magnetograms.
This AR mask, shown with orange contour in Figure~\ref{fig:masks}, highlights areas with magnetic field strength of over $30~\si{G}$ (see Fig.~\ref{fig:sdo_mag}) to ensure that only stronger than quiet-Sun fields are associated with the AR. 
This mask is then applied to the corresponding AIA observations (for example, see Fig.~\ref{fig:sdo_photosphere} for its application in the 1600\,\AA\ data) to isolate magnetically active areas relevant to our study through visual inspection.
The value of $30~\si{G}$ was empirically chosen so as to separate quiet-Sun from active-region fields.

\begin{figure*}
    \centering
    \begin{tabular}{cM{\textwidth/6}M{\textwidth/6}M{\textwidth/6}M{\textwidth/6}M{\textwidth/6}}
        \toprule
        \multicolumn{6}{c}{\textbf{AR 11429}}\\
        \midrule
         & 94\AA & 131\AA & 171\AA & 193\AA & 1600\AA \\
        \midrule
        AIA & \includegraphics[width=\linewidth]{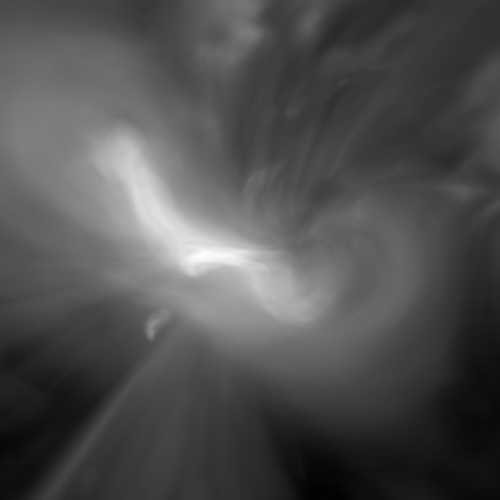} & \includegraphics[width=\linewidth]{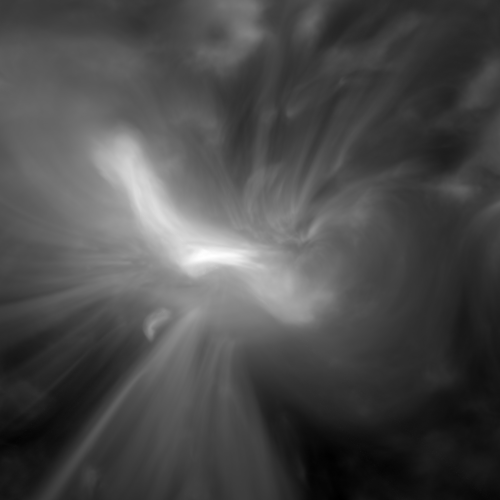} & \includegraphics[width=\linewidth]{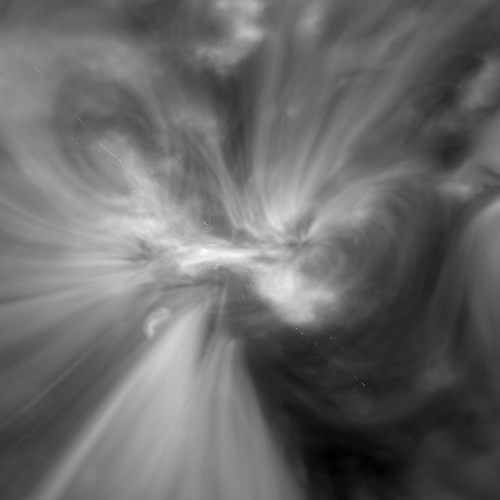} & \includegraphics[width=\linewidth]{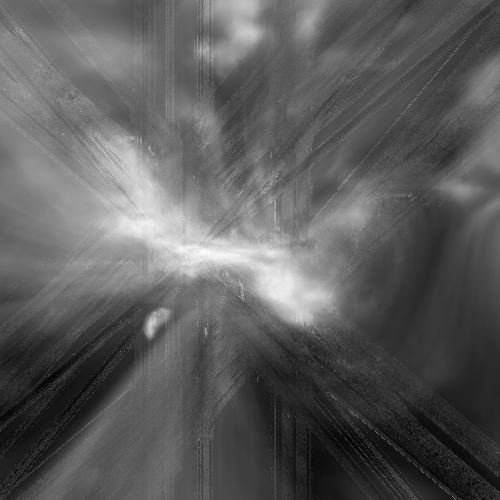} & \includegraphics[width=\linewidth]{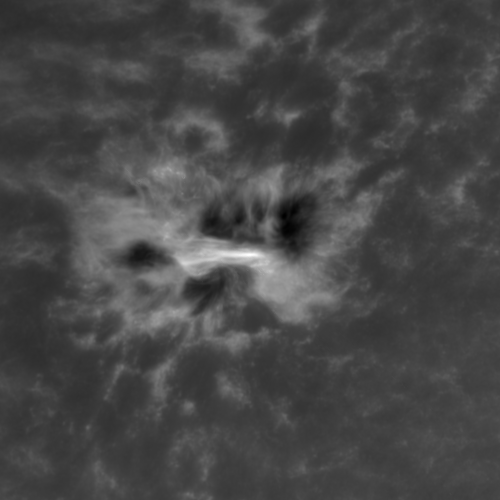} \\
        Mask & \includegraphics[width=\linewidth]{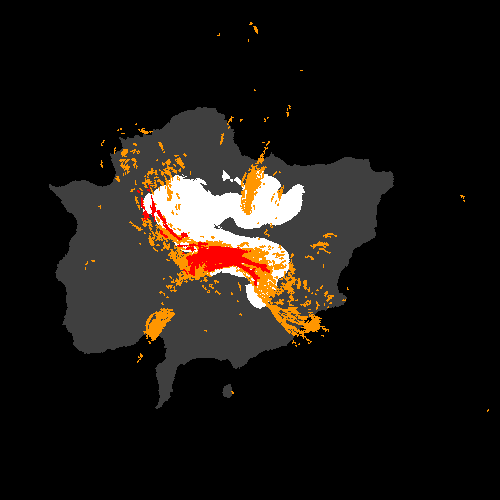} & \includegraphics[width=\linewidth]{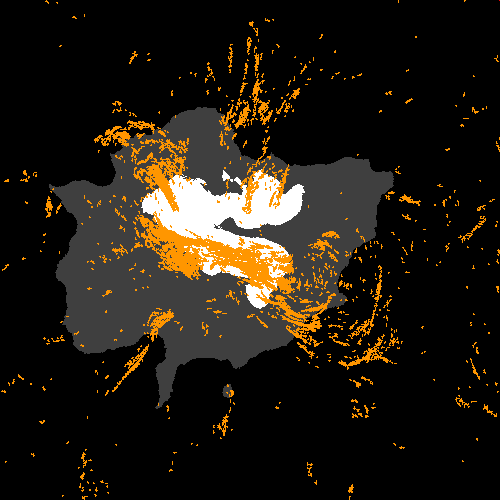} & \includegraphics[width=\linewidth]{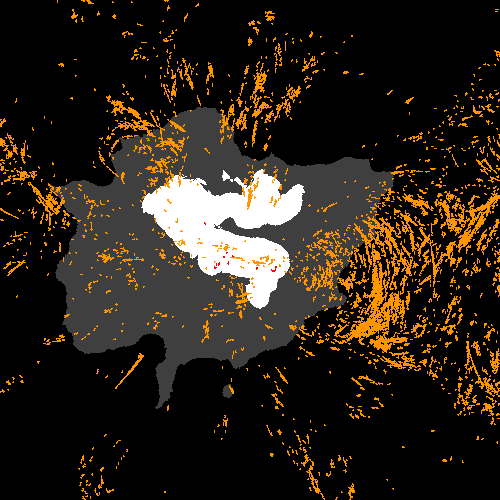} & \includegraphics[width=\linewidth]{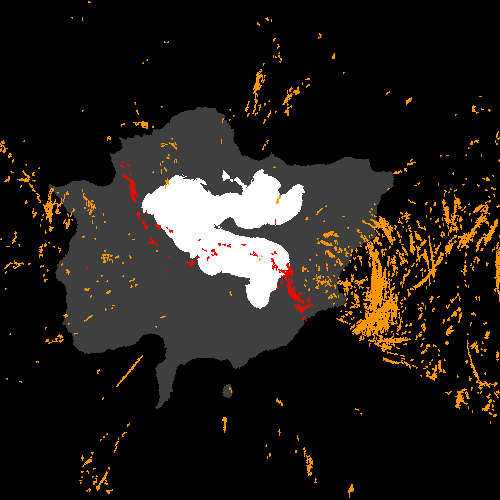} & \includegraphics[width=\linewidth]{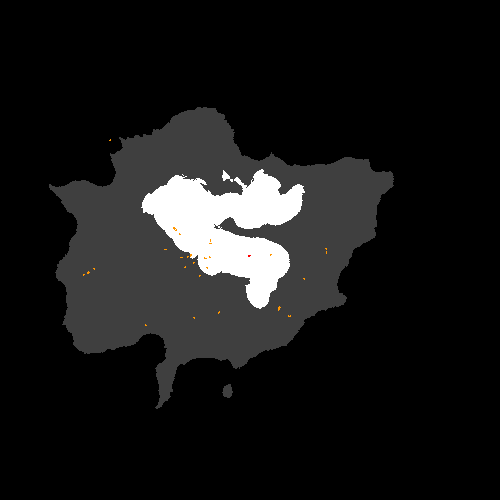} \\
        \midrule
        \multicolumn{6}{c}{\textbf{AR 13186}}\\
        \midrule
        AIA & \includegraphics[width=\linewidth]{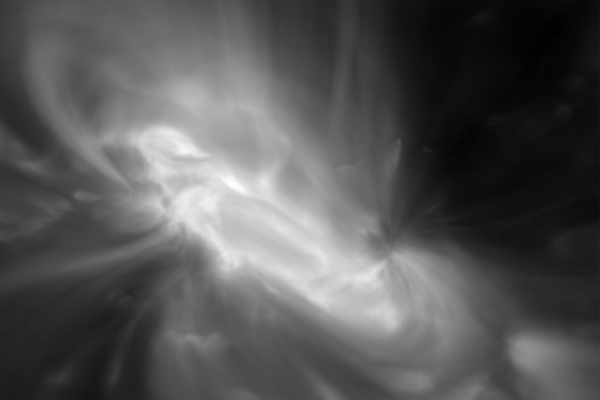} & \includegraphics[width=\linewidth]{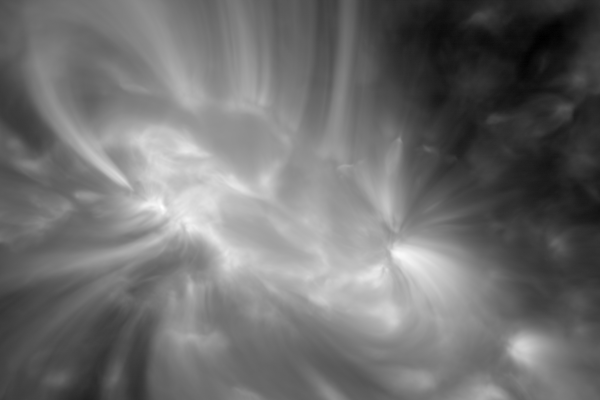} & \includegraphics[width=\linewidth]{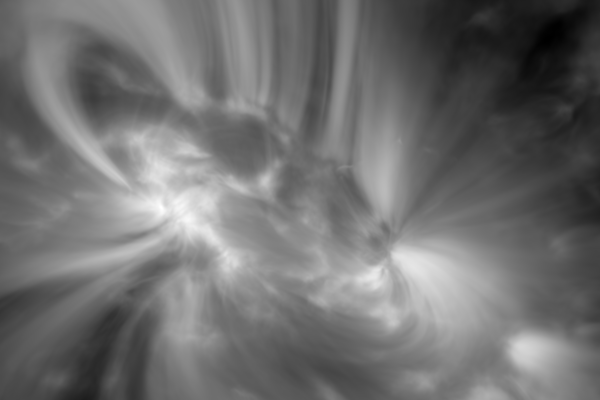} & \includegraphics[width=\linewidth]{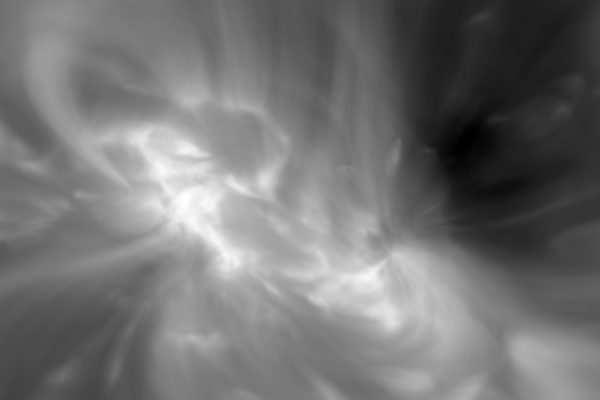} & \includegraphics[width=\linewidth]{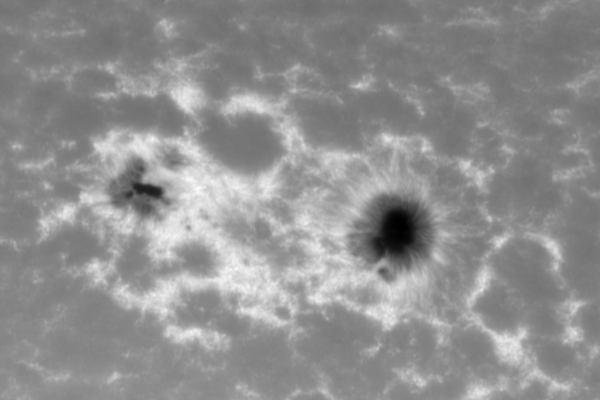} \\
        Mask & \includegraphics[width=\linewidth]{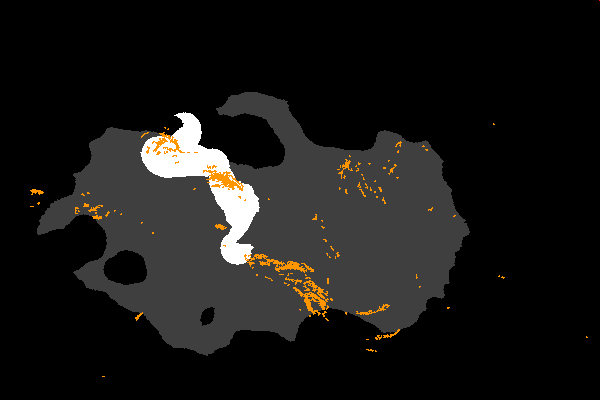} & \includegraphics[width=\linewidth]{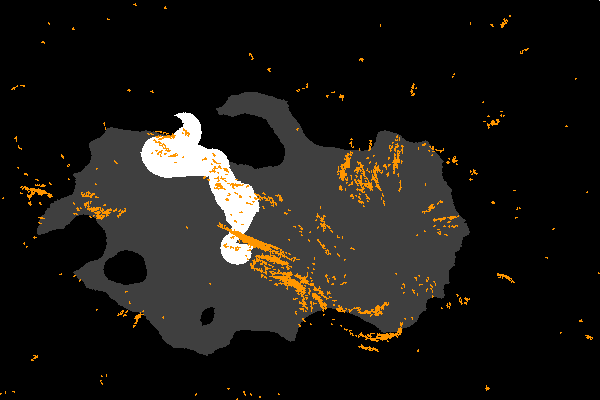} & \includegraphics[width=\linewidth]{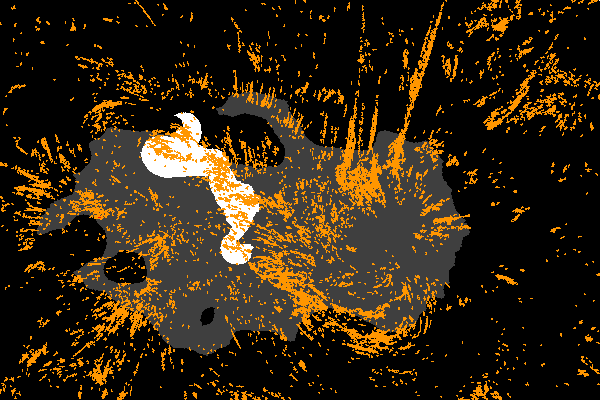} & \includegraphics[width=\linewidth]{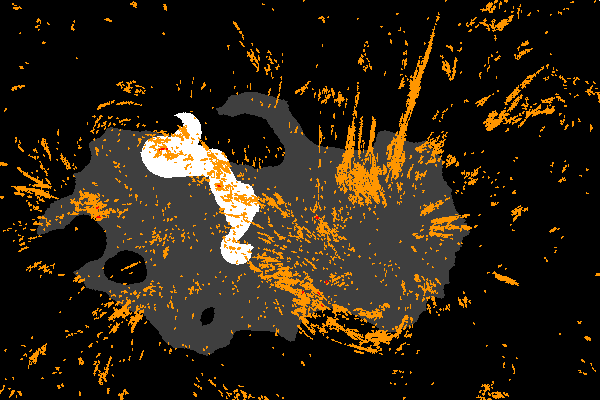} & \includegraphics[width=\linewidth]{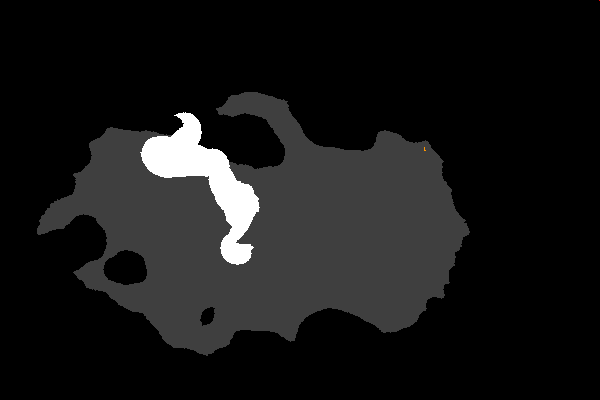} \\
        \bottomrule
    \end{tabular}
    \caption{Multiple-wavelength observations of the flare-imminent (AR 11429 on top) and the non-flaring (AR 13186 in the bottom) configurations.
    Each panel shows observations made on the full 24-hour set.
    The colour scale for the masks shows the AR mask (grey), the PIA (white), the 3-$\sigma$ (orange) and the power law (red) TBs.}
    \label{tab:multi wavelengths}
\end{figure*}

By combining masks of the detected 3-$\sigma$ and power law threshold brightenings with the PIA observations and the AR mask, the resulting images are shown in Figure~\ref{tab:multi wavelengths}.
This figure presents a combined view of multi-wavelength observations for both the flare-imminent phase of AR 11429 and a random, non-flaring phase of AR 13186, in terms of TBs and their spatial relation to the PILs.

We note that the 3-$\sigma$ brightenings exhibit a non-uniform distribution across multiple channels.
More specifically, in Figure~\ref{tab:multi wavelengths}, and throughout the entire sequence, we observe that the 3-$\sigma$ brightenings show a better correspondence with the PIL, primarily in 94\,\AA.
In contrast, the correlation is weaker in channels such as 171\,\AA, 193\,\AA, and 1600\,\AA, where the 3-$\sigma$ TBs are more spread out, extending beyond the active region core into its periphery.
The 1600\,\AA\ channel stands out in particular, as the 3-$\sigma$ TBs detected are both very scarce and small.

The 131\,\AA\ channel exhibits a combination of the distribution patterns observed in the 94\,\AA\ and 171\,\AA\ channels.
This is expected, as the regions it probes (transition region and flaring corona) and the associated temperatures (log($T$) of 5.6 and 7.0) share characteristics with those seen in both the 94\,\AA\ and 171\,\AA\ channels \citep[see Table 1 in][]{Lemen2012}.

Additionally, after processing our data, in addition to the higher concentration of 3-$\sigma$ TBs around the PIL with the 94\,\AA\ channel, power law TBs appear more consistently in this channel too (see Fig.~\ref{tab:multi wavelengths}), which is hereby the data that we choose to perform our analysis.
Let us now focus on the 94\,\AA\ channel to evaluate how the previously introduced 3-$\sigma$ and power law thresholds function as diagnostic tools for identifying active regions capable of producing X-class flares, which are often associated with CMEs.
Notably, \citet{Leka2023} also found that the 94\,\AA\ filter yields the largest number of parameters with strong discriminating power.

\section{Testing the brightening detection method}\label{sec:test}

In this section, we test whether our BADPIT method can identify notable differences within a 24-hour window between the quiescent, non-flaring AR 13186 and the highly dynamic, flaring AR 11429.
Our goal is to analyse AR behaviour prior to and leading up to significant X-class flaring events.
For AR 13186, we randomly selected a 24-hour interval of flare-quiet activity when the AR was located near the disk centre.
For AR 11429, we select a period free of X-class flares immediately before two such flares occurred, starting from 00:02 on March 7, 2012.
Therefore, we analyse the 24-hour period beginning at 00:00 on March 6, 2012.
The chosen interval for the flaring AR is not free of M-class flares that are sometimes also considered as major flares, however in this case study our focus is on the differences between a flare-quiet configuration and an X-class flare-imminent one.
A more comprehensive analysis is currently underway and will examine these differences over a broader and finer flare classification scale than the exploratory case study presented here.

The periods analysed for these two ARs are presented in Table~\ref{tab:datasets}, and the number of flares detected in each GOES class for both ARs during the observation periods is shown in Table~\ref{tab:flare classes}.

\begin{table*}
    \centering
    \begin{tabular}{||c|c|c||}
        \hhline{|t:===:t|}
        \textbf{GOES} & \multicolumn{2}{c||}{\textbf{Reported flares}}\\
        \hhline{||~|--||}
        \textbf{classification} & \textbf{13186} & \textbf{11429} \\
        \hhline{|:===:|}
        C & 0 & 5 \\
        M & 0 & 5 \\
        X & 0 & 0 \\
        \hhline{|b:===:b|}
    \end{tabular}
    \caption{Number of flares detected in every GOES-class category for both ARs during the period of observation from GOES flare catalogue.}
    \label{tab:flare classes}
\end{table*}

\begin{figure*}
   \centering
   \begin{tabular}{M{\textwidth/7}M{\textwidth/6}M{\textwidth/6}M{\textwidth/6}M{\textwidth/6}}
        \toprule
        \multicolumn{5}{c}{\textbf{AR 11429}}\\
        \midrule
        Time (UT) & \textbf{\shortstack{$00:00$ \\ $\rightarrow$ \\ $05:59$}} & \textbf{\shortstack{$06:00$ \\ $\rightarrow$ \\ $11:59$}} & \textbf{\shortstack{$12:00$ \\ $\rightarrow$ \\ $17:59$}} & \textbf{\shortstack{$18:00$ \\ $\rightarrow$ \\ $23:59$}} \\
        \midrule
        94\AA\, AIA average image & \includegraphics[width=\linewidth]{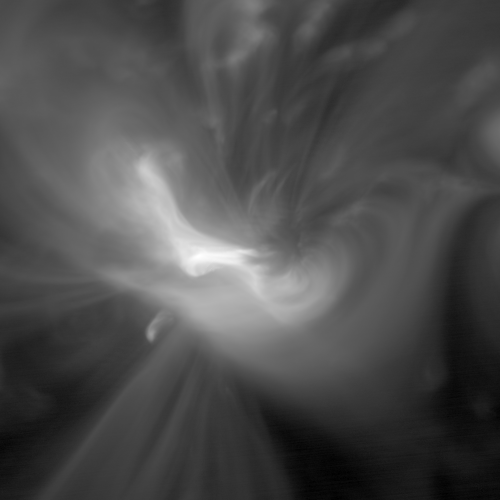} & \includegraphics[width=\linewidth]{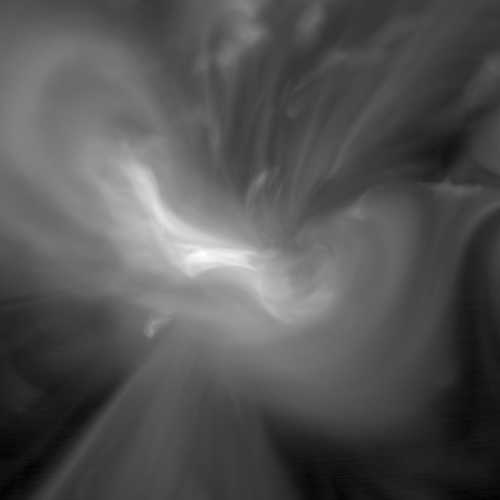} & \includegraphics[width=\linewidth]{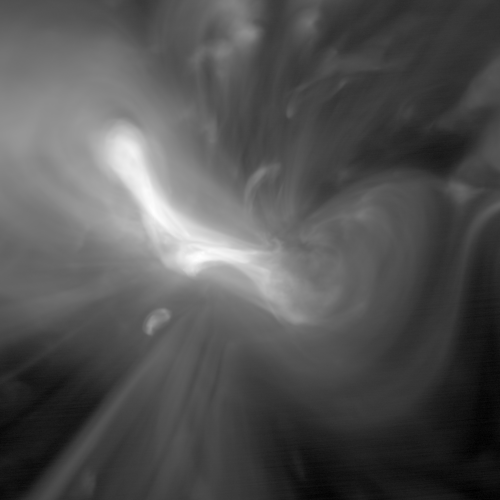} & \includegraphics[width=\linewidth]{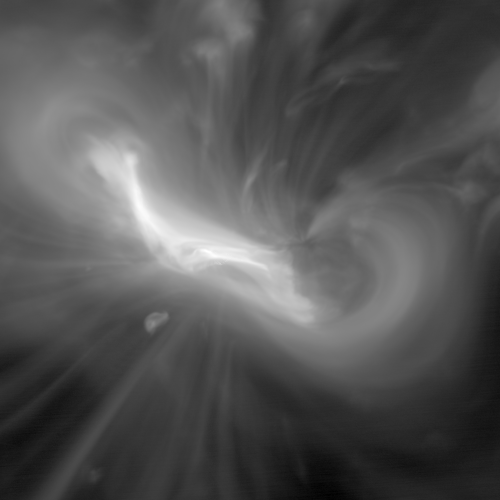} \\
        TBs and PIA & \includegraphics[width=\linewidth]{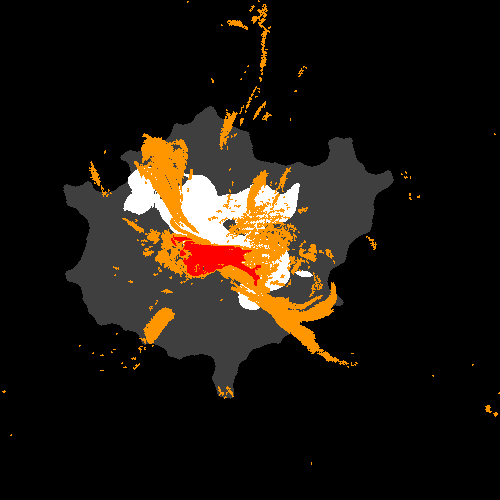} & \includegraphics[width=\linewidth]{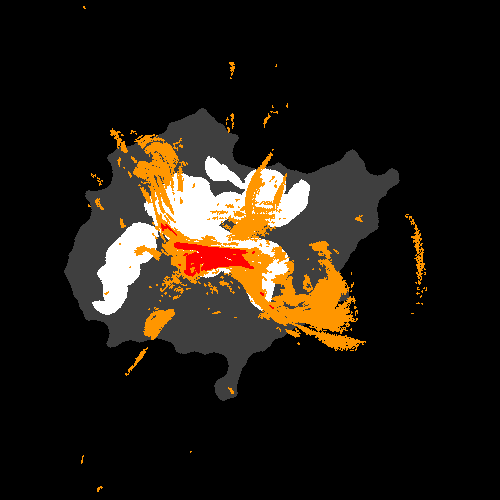} & \includegraphics[width=\linewidth]{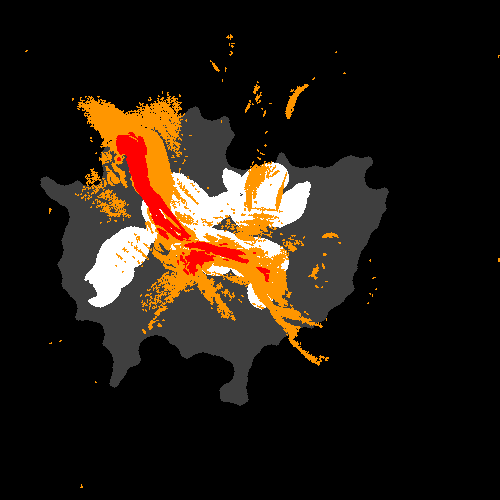} & \includegraphics[width=\linewidth]{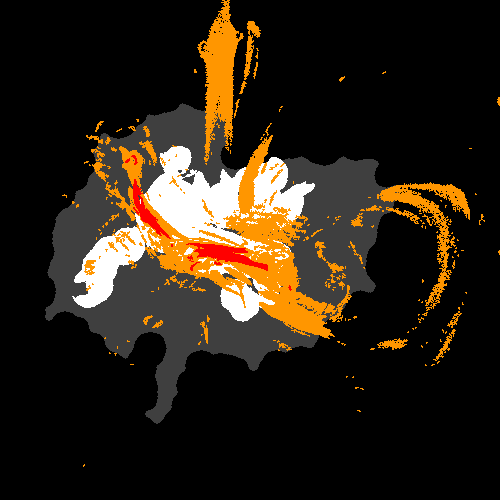} \\
        \midrule
        \multicolumn{5}{c}{\textbf{AR 13186}}\\
        \midrule
        Time (UT) & \textbf{\shortstack{$00:00$ \\ $\rightarrow$ \\ $05:59$}} & \textbf{\shortstack{$06:00$ \\ $\rightarrow$ \\ $11:59$}} & \textbf{\shortstack{$12:00$ \\ $\rightarrow$ \\ $17:59$}} & \textbf{\shortstack{$18:00$ \\ $\rightarrow$ \\ $23:59$}} \\
        \midrule
        94\AA\, AIA average image & \includegraphics[width=\linewidth]{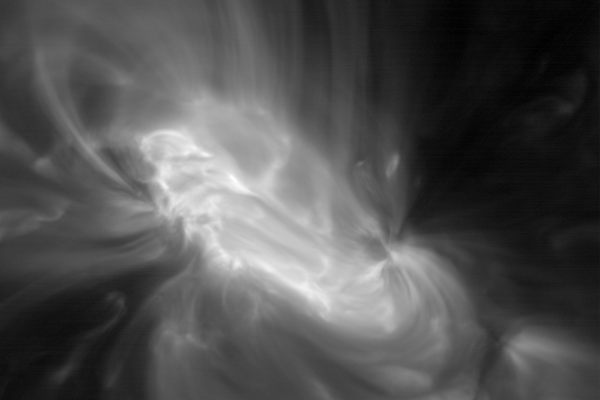} & \includegraphics[width=\linewidth]{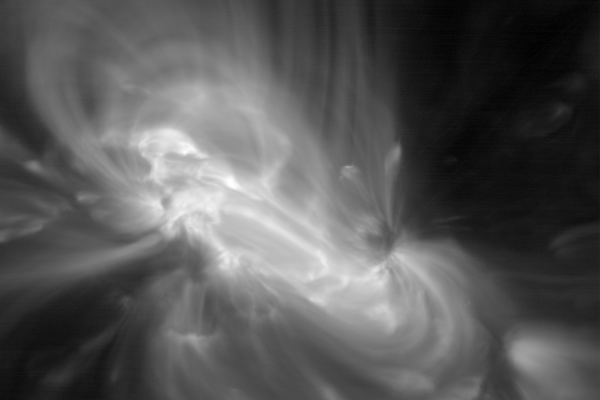} & \includegraphics[width=\linewidth]{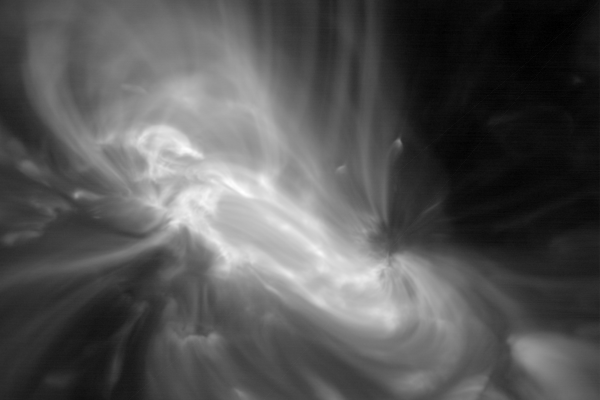} & \includegraphics[width=\linewidth]{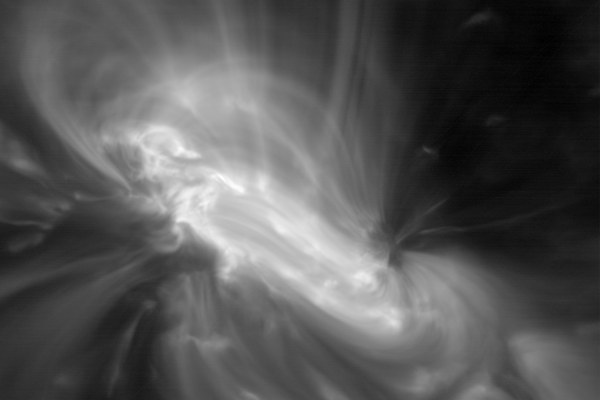} \\
        TBs and PIA & \includegraphics[width=\linewidth]{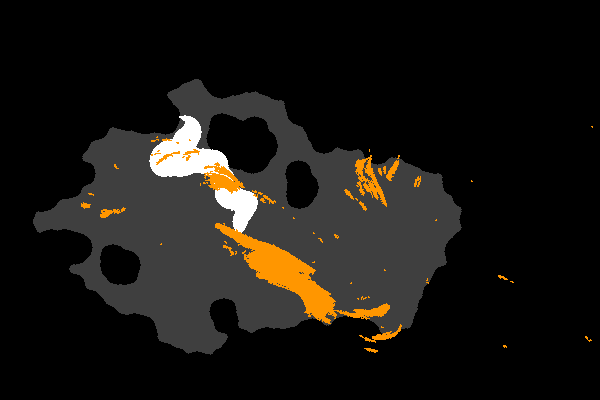} & \includegraphics[width=\linewidth]{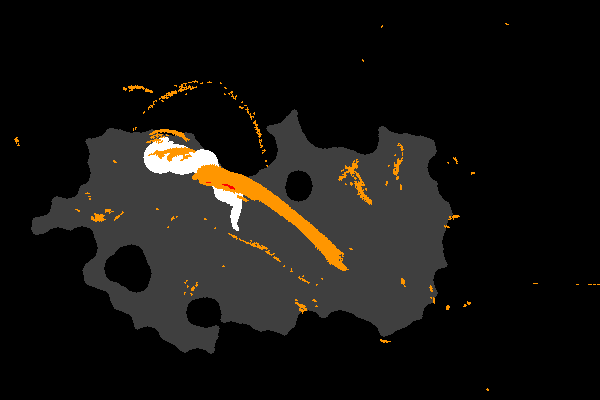} & \includegraphics[width=\linewidth]{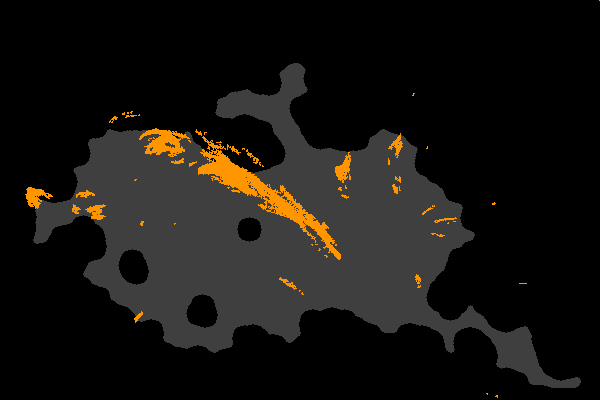} & \includegraphics[width=\linewidth]{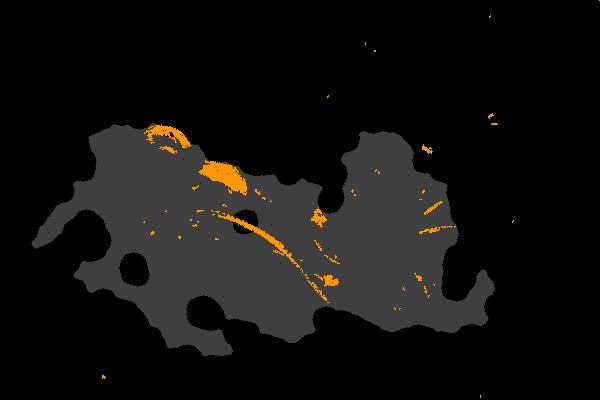} \\
        \bottomrule
    \end{tabular}
    \caption{Evolution of the two ARs through the 24 hours of observation and the brightenings distribution associated with every time sequence.
    The location of the PIA is indicated by white pixels.
    The orange pixels show the brightenings detected through the 3-$\sigma$ threshold method, while in red are the ones from the power law threshold method.
    Every panel here shows the activity corresponding of 6-hour intervals.}
    \label{fig:timeserie}
\end{figure*}

We apply the BADPIT method, as described in Section~\ref{methodo}, to the two active regions, and the results are shown in Figure~\ref{fig:timeserie}, where the distribution of detected brightenings is presented for four consecutive 6-hour intervals.
In Figure~\ref{fig:timeserie} masks, the AR is indicated in grey and the PIA in white.
The orange pixels represent brightenings detected using the 3-$\sigma$ threshold method, while those detected with the power law threshold method appear in red.

\begin{figure*}
    \centering
    \subfloat[AR 13186]{\includegraphics[width=0.45\linewidth]{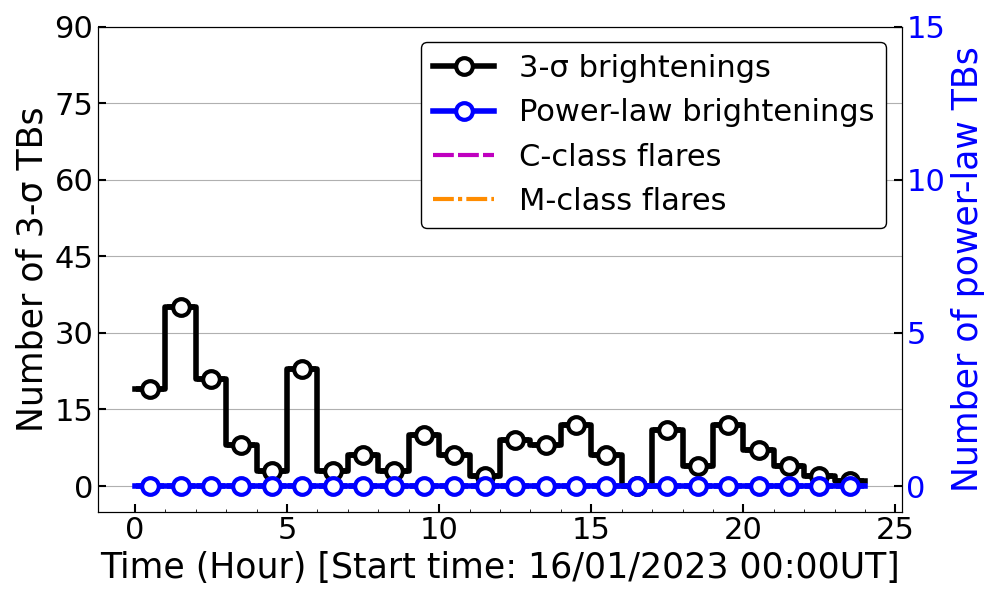}\label{nb_13186}}
    \hspace{0.05\linewidth}
    \subfloat[AR 11429]{\includegraphics[width=0.45\linewidth]{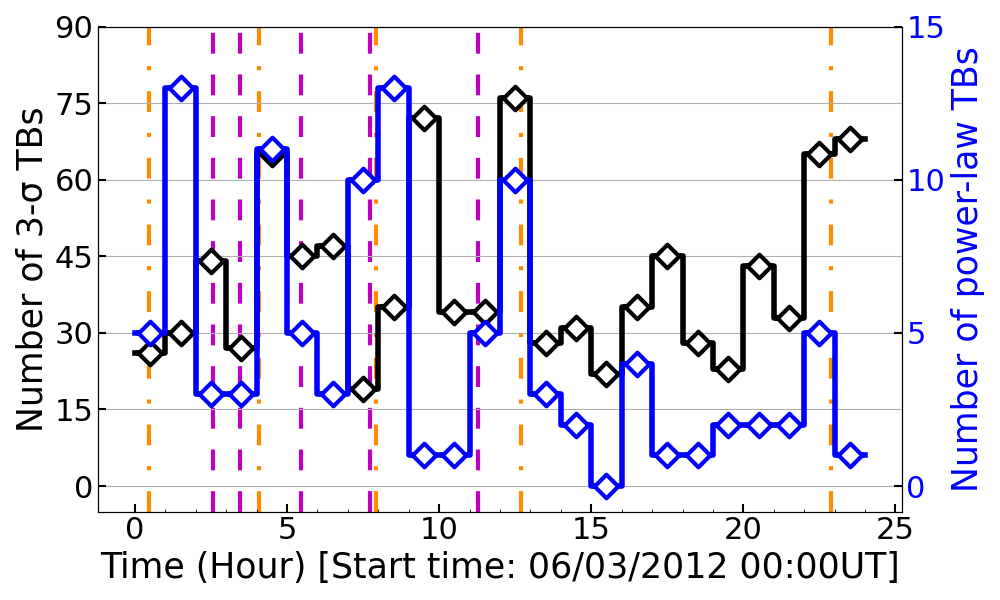}\label{nb_11429}}
    
    \caption{Number of brightenings detected using the 3-$\sigma$ (black) and the power law (blue) threshold methods in the 94\,\AA\ channel.
    Panel a shows the results obtained for AR 13186 (circle-shaped marker), and Panel b for AR 11429 (diamond-shaped marker).
    The orange and red vertical dashed lines correspond, respectively, to the starting times of C- and M-class flares that occurred in the ARs.
    The TBs were counted over timeseries of 1-hour intervals.}
    \label{fig:nb_timeserie}
\end{figure*}

The next step is to evaluate how the two threshold levels function as diagnostic tools for identifying active regions capable of producing large flares.
The evolution of the number of detected brightenings, analysed using both the 3-$\sigma$ and power law threshold methods, is presented for the two ARs in Figure~\ref{fig:nb_timeserie}.

Using both the 3-$\sigma$ and power law threshold methods we generally detect more TBs in AR 11429 (flaring) compared to AR 13186 (non-flaring).
We find around five times as many brightenings detected with the 3-$\sigma$ method.
The power law brightenings show an even more drastic difference with no brightenings detected in the non-flaring AR while an average of 4 TBs per hour are detected for the flaring AR.
We reiterate here that the majority of TBs counted in Figure~\ref{fig:nb_timeserie} are distributed over the PIA.

With the way our detection method works, the TBs detected include potential sub-flaring events but also up to C- and M-class flares and does not distinguish them.
However, Figure~\ref{nb_11429} shows no dramatic increase in the number of brightenings for time series including both C- and M-class flares: they are always within the same order of magnitude with the numbers of TBs detected in other intervals.
This is expected, as even large flare-related events are detected as only a limited number of sources, depending on the flare morphology, whereas the total event population is dominated by the much more numerous smaller-scale brightenings.
In Figure~\ref{nb_13186}, we observe a consistently lower number of TBs and an overall absence of high-intensity brightenings detected with the power law method during the observation period.

In summary, both detection methods reveal a clear difference in the number of TBs between the flaring and the non-flaring ARs.
However, the power law threshold method, which targets higher-intensity brightenings, shows stronger distinguishing power, with no TBs detected for the non-flaring AR.
The presence of brightenings with distributions obeying power laws should not be taken as a binary indicator of an incoming flare.
However, a significant difference in the number of such TBs should be expected and looks promising, with a five times difference in our case study.

\begin{figure*}
    \centering
    \subfloat[Intensity]{\includegraphics[width=0.45\linewidth]{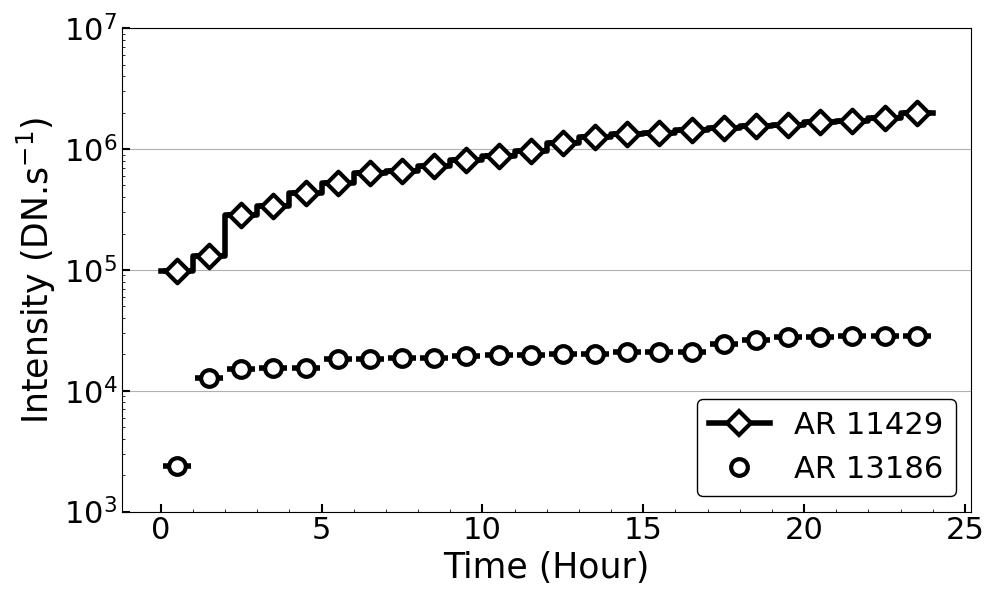}\label{cumul_sig}}
    \hspace{0.05\linewidth}
    \subfloat[Unsigned magnetic flux]{\includegraphics[width=0.45\linewidth]{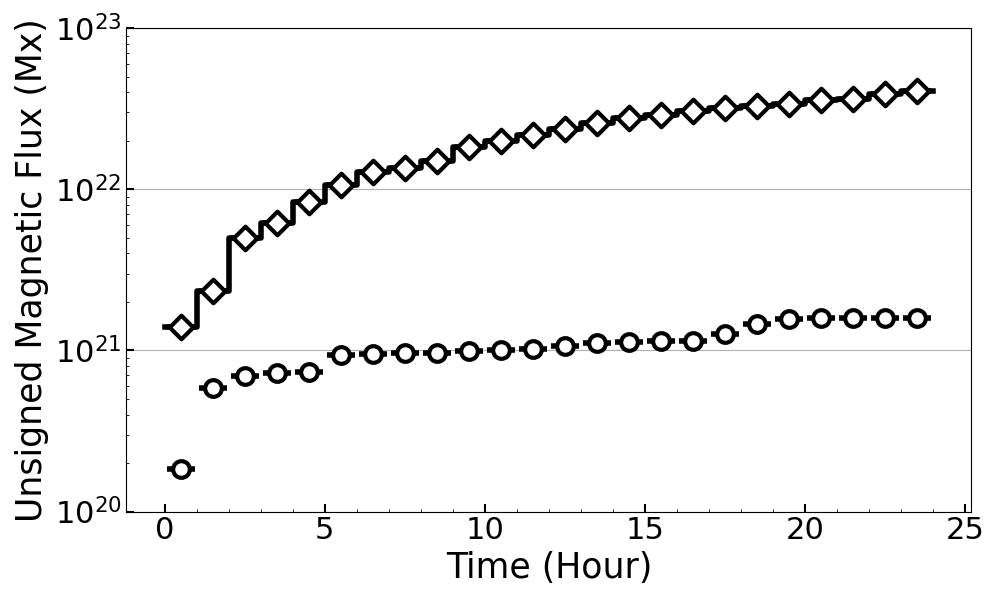}\label{cumul_mag}}
    
    \caption{Cumulative evolution of the total 94\,\AA\ intensity (panel a) and unsigned magnetic flux (panel b) of the 3-$\sigma$ brightenings detected in AR 13186 (non-flaring; circle dash marker) and 11429 (flaring; full line and diamond marker).
    The measure is done over timeseries of 1-hour intervals.}
    \label{fig:int_timeserie}
\end{figure*}

Next, we focus on the cumulative evolution of the TBs' intensity,  shown in Figure~\ref{fig:int_timeserie}.
The cumulative evolution of intensity was measured for the 3-$\sigma$ TBs (see Fig.~\ref{cumul_sig}).
It is then clear that the cumulative intensity of the non-flaring AR is significantly lower than that of the flaring AR, with approximately two orders of magnitude difference when using the 3-$\sigma$ threshold.
We made analogous plots for the cumulative magnetic flux co-located to the TBs, which exhibited the same overall trends, differing only in the $y$-axis scales (see Fig.~\ref{cumul_mag}).

The cumulative intensity steadily increases for the flaring AR during the 24-hour study period.
In contrast, for the non-flaring AR, as shown in Figure~\ref{fig:int_timeserie}, the cumulative intensity remains nearly constant throughout the 24-hour period.
A slight increase in cumulative intensity is observed, but it is not nearly as significant as that of the flaring AR.

All the previous results were obtained with timeseries intervals of one hour but we repeated the analysis for intervals of 2, 3, 4 and 6 hours.
As it was natural we observed differences between the flaring and the non-flaring ARs in both number of brightenings (Fig.~\ref{fig:nb_timeserie}) and cumulative intensities (Fig.~\ref{fig:int_timeserie}).
The evolution of the cumulative intensity stayed consistent and was almost not affected by the changes in the sampling interval of the timeseries.
The number of TBs detected in time series constructed over 1-hour intervals display many fluctuations correlated with reported events.
Increasing the interval length leads to a reduction of these fluctuations and a harmonisation of the results.
This implies that the number of brightenings is more affected by short-term activity than the cumulative brightness. 

\begin{figure*}
    \centering
    \subfloat[AR 13186]{\includegraphics[width=0.45\linewidth]{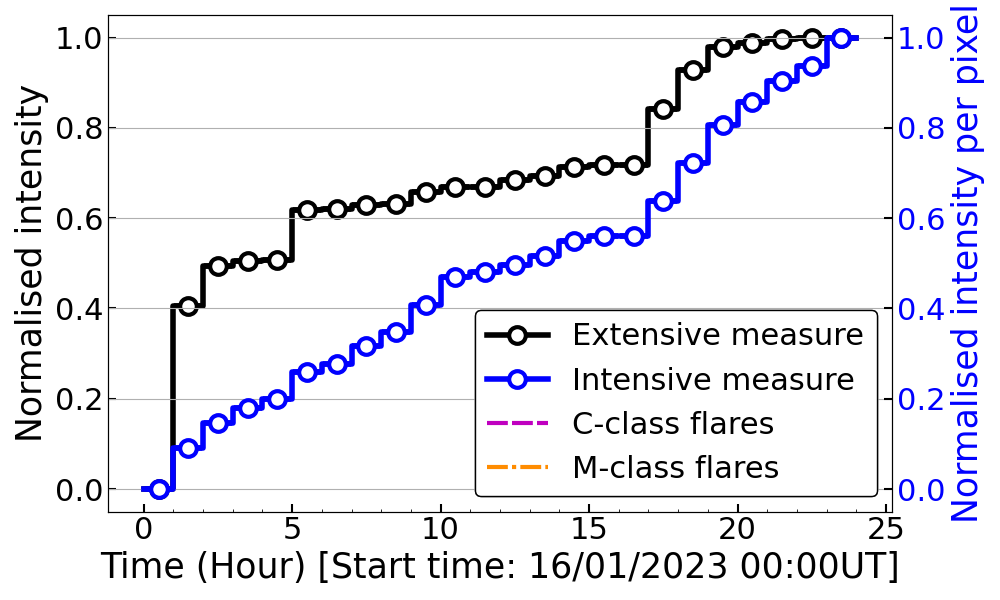}\label{int-ext 13186}}
    \hspace{0.05\linewidth}
    \subfloat[AR 11429]{\includegraphics[width=0.45\linewidth]{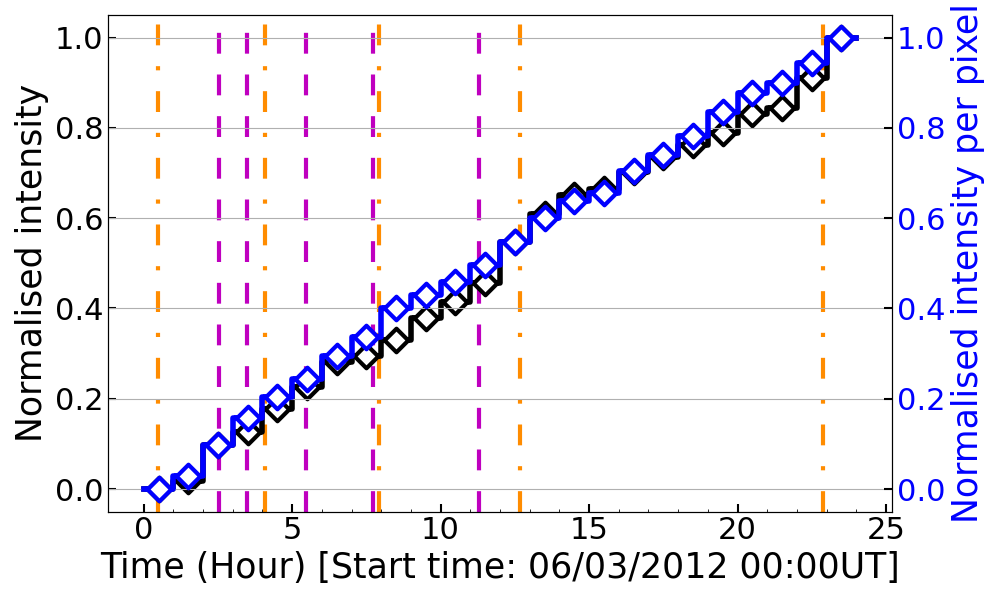}}
    \caption{Comparison of the extensive (cumulative intensity; black) and intensive (cumulative intensity per pixel; blue) measures of the intensity associated with the 3-$\sigma$ brightenings detected through 94\,\AA\ observations of AR 13186 (circle marker) on the left panel (a) and AR 11429 (diamond marker) on the right one (b). Here, we use a sampling interval of 1 hour.}
    \label{fig:int-ext}
\end{figure*}

\citet{welsch_what_2009} characterised active region properties as extensive when they scale with active region size, and intensive when they do not directly depend on it.
Following this classification, we associate extensive parameters with population sums and intensive parameters with population means.
It has been suggested that extensive parameters are more relevant to reflect flaring behaviour under 24-hour periods (the result might be different with 48-hour periods) \citep[and references therein]{bobra_solar_2015, bobra_predicting_2016}.
Our results, comparing the extensive measure of intensity (i.e., the sum of the intensity over the brightenings) and its corresponding intensive measure (i.e., the mean intensity of the brightenings), are shown in Fig.~\ref{fig:int-ext}.

This said, the evolution patterns of the two measures shown in their normalised form in Fig.~\ref{fig:int-ext} reveal distinct behaviours.
For the flaring active region, both extensive and intensive measures show a very similar linear trend.
In contrast, the non-flaring active region exhibits a significant step-like, nonlinear evolution that is more notable in the extensive measure.
This step-like behaviour is related to minor activity: for example, in AR 13186, a sudden increase in intensity occurs shortly after 2023-01-16 02:00, 06:00 and 18:00.
These jumps are clearly visible in the extensive measure but not so much in the intensive one.
The likely explanation, supported by the inspection of related AIA movies, is that the surface area of the brightenings increased, which diluted the average intensity and smoothed out the response in the intensive metric.
Since the intensive value represents the mean intensity over the brightening area, the peak observed in the extensive curve becomes less pronounced.
This difference in evolutionary patterns seems aligned with the previous finding (see above for references) that extensive parameters are better suited to observe flaring behaviour.

\section{Conclusion}\label{sec:discu}
We have developed the Brightenings AnD Polarity Inversion Tracking (BADPIT) method to detect and monitor TBs.
TBs are short-lived, localised brightenings that appear significantly more intense than their immediate surroundings.
In the literature, TBs are considered to potentially represent early signatures of the build-up toward major flares, as they tend to form clusters near the locations of future flare ribbons, typically concentrated along strong-gradient PILs and in regions of enhanced magnetic energy density \citep[and references therein]{dissauer_brightenings_2025}.

In this work, we demonstrate that BADPIT provides a promising approach for detecting TBs in a pair of magnetically complex active regions with contrasting activity: one of them produced major flares while the other was quiet.
To implement the BADPIT methodology, we developed a new algorithm that successfully desaturates AIA images.
This method may also be useful for other studies using SDO/AIA data, particularly for observation taken in the 94\,\AA, 131\,\AA, 171\,\AA, 193\,\AA, and 1600\,\AA\ channels.

To focus on the study of TBs around the PIL, we use the 94,\AA\ channel, as this channel shows a higher concentration of TB clusters near the PIL than any other channel for both TB categories tracked in this study.
This is consistent with the findings of \citet{Leka2023}, who reported that the 94\,\AA\ AIA EUV channel contains the largest number of parameters with strong flaring discriminant power.

The BADPIT method tracks TB activity across multiple SDO/AIA EUV channels using two threshold methods:
\begin{itemize}
\item 3-$\sigma$ threshold method: A pixel-based threshold defined as three times the standard deviation above the mean of a given pixel’s light curve.
This method assumes the intensity signal follows a Gaussian distribution, such that 99\% of values fall within the 3-$\sigma$ interval, making deviations statistically significant.
    
\item power law threshold method: A global thresholding approach that identifies only high-intensity brightenings, using a spatially and temporally invariant threshold independent of the pixel’s background intensity.
The threshold is defined as the separation value at which the intensity histogram deviates from the power law fitted distribution.
\end{itemize}

To investigate the TB activity within the strong magnetic gradient area of the ARs, we followed the method described by \citet{schrijver_characteristic_2007} to identify the PIA using the corresponding LoS SDO/HMI magnetograms (see Fig.~\ref{fig:timeserie}).
We applied the BADPIT methodology to two ARs with distinctly different levels of flaring activity: AR 11429, which produced several strong flares, and AR 13186, which produced none (see Table~\ref{tab:flare classes}).
We refer to these two ARs as the flaring and non-flaring ARs, respectively.
For each region, we analysed a continuous 24-hour observation window.
In the case of the flaring region (AR 11429), the 24-hour window was selected to cover the period leading up to a couple of X-class flares.
Based on this analysis, we found the following:

\begin{itemize}
\item The 94\,\AA\ channel proved to be the most suitable channel for the tracking (using both the 3-$\sigma$ and power law methods) of TBs concentrated around the PIA, in quantities sufficient for their unambiguous detection.
However, TBs do appear in other AIA channels as well.
The 3-$\sigma$ brightenings follow distinct spatial patterns adhering in each channel.
They appear very small and scarcely scattered across the active region in the 1600\,\AA\ observations, follow loop-like structures in the 171\,\AA\ and 193\,\AA\ channels, and are concentrated within the PIA in the 94\,\AA\ channel.

\item The 94\,\AA\ channel shows the strongest concentration of 3-$\sigma$ brightenings within the PIA, as well as a non-negligible number of power law TBs (an average of 4 TBs per hour for the flaring AR) concentrated above the PIA.
We deemed that 94\,\AA\ observations are the most relevant for identifying TBs associated with strong flares.
    
\item When counting the number of TBs, we identify approximately five times as many in the flaring active region compared to the non-flaring one, using the 3-$\sigma$ threshold method on 94\,\AA\ data (see Fig.~\ref{fig:nb_timeserie}).
Furthermore, the power law threshold method demonstrates even stronger discriminating power in separating TB activity between the two regions, based on the number of detected TBs with no TBs detected in the non-flaring configuration (see Fig.~\ref{fig:nb_timeserie}).

\item We also examined the cumulative evolution of brightening intensity over the 24-hour analysis windows, based on 94\,\AA\ data.
This comparison clearly differentiates the two active regions, revealing an order-of-magnitude difference in cumulative intensity when applying the 3-$\sigma$ method (Fig.\ref{fig:int_timeserie}).
\end{itemize}

Overall, we find that BADPIT, which combines image desaturation, dual thresholding methods, and comparison with the PIA, shows promising potential as a tool to investigate the evolution of TBs prior to flaring events.
However, the size of ARs we studied is very limited, therefore our work should be  regarded as a proof-of-concept/pilot study.
BADPIT's reliability as a forecast tool will be assessed in a follow-up study in which a large sample of ARs will be analysed, which will allow us to derive statistically significant conclusions.
Future efforts will also focus on refining the power law threshold level to account for the gradual decrease in AIA sensitivity as the instrument continues to age.

\begin{acks}[Acknowledgements]
We wish to thank the anonymous reviewer whose comments helped us meaningfully improve and consolidate the manuscript.
This work is part of the SWATNet project funded by the European Union’s Horizon 2020 research and innovation programme under the Marie Skłodowska-Curie grant agreement No 955620.

MBK is grateful for the Leverhulme Trust Found ECF-2023-271.  MBK and RE acknowledge the NKFIH OTKA (Hungary, grant No. K142987).
R.E. is grateful to Science and Technology Facilities Council (STFC, grant No. ST/M000826/1) UK, acknowledges  PIFI (China, grant number No. 2024PVA0043) and the NKFIH (Hungary)  Excellence Grant (grant nr TKP2021-NKTA-64) for enabling this research. MBK and RE acknowledge the NKFIH OTKA (Hungary, grant No. K142987) and the ESA COMMERCIAL APPLICATIONS OF SPACE WEATHER DATA FS - EXPRO+1-12676 2025. A.N. and S.P. acknowledge support by the ERC Synergy Grant (GAN: 810218) "The Whole Sun".
\end{acks}


\bibliographystyle{spr-mp-sola}
\bibliography{biblio}  

\end{document}